# *Astrobiology*



## Spectral signatures of photosynthesis I: Review of Earth organisms

| | |
|---|---|
| Journal: | *Astrobiology* |
| Manuscript ID: | AST-2006-0105 |
| Manuscript Type: | Research Articles (Papers) |
| Date Submitted by the Author: | 22-Nov-2006 |
| Complete List of Authors: | Kiang, Nancy; NASA Goddard Insitute for Space Studies; CalTech, Infrared Processing and Analysis Center (IPAC)<br>Siefert, Janet; Rice University, Dept. of Statistics<br>Govindjee, Govindjee; University of Illinois at Urbana-Champaign, Departments of Plant Biology and Biochemistry<br>Blankenship, Robert; Washington University, Department of Biology and Chemistry<br>Meadows, Victoria; California Institute of Technology, Spitzer Science Center |
| Keyword: | Anoxygenic Photosynthesis, Oxygenic Photosynthesis, Astrobiology, Spectroscopic Biosignatures, Red Edge |
| | |

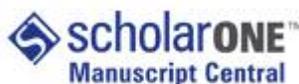





**Short title:**
**Spectral signatures of photosynthesis:  Earth review**

**Full title:**
**Spectral signatures of photosynthesis I:  Review of Earth organisms**


*Nancy Y. Kiang[1,6], Janet Siefert[2,6], Govindjee[3], Robert E. Blankenship,[4] Victoria S. Meadows.[2,6]

[1]NASA Goddard Institute for Space Studies, U.S.A.
[2]Dept. of Statistics, Rice University
[3]Departments of Plant Biology and Biochemistry, University of Illinois at Urbana-Champaign, U.S.A.
[4]Department of Biology and Chemistry, Washington University, U.S.A.
[5]Spitzer Science Center, California Institute of Technology, USA
[6]NASA Astrobiology Institute

*To whom correspondence should be addressed:  Nancy Y. Kiang, NASA Goddard Institute for Space Studies, New York, NY  10025, fax1: (626) 568-0673, fax2: (212) 678-5552, tel1: (626) 395-1815, tel2: (212) 678-5587, cell: (949) 439-3416, email: nkiang@giss.nasa.gov






# DRAFT, Kiang, Astrobiology

## Abstract


Why do plants reflect in the green and have a "red edge" in the red, and should extrasolar photosynthesis be the same? We provide: 1) a brief review of how photosynthesis works; 2) an overview of the diversity of photosynthetic organisms, their light harvesting systems, and environmental ranges; 3) a synthesis of photosynthetic surface spectral signatures; 4) evolutionary rationales for photosynthetic surface reflectance spectra with regard to utilization of photon energy and the planetary light environment. We found the "NIR end" of the red edge to trend from blue-shifted to reddest for (in order): snow algae, temperature algae, lichens, mosses, aquatic plants, and finally terrestrial vascular plants. The red edge is weak or sloping in lichens. Purple bacteria exhibit possibly a sloping edge in the NIR. More studies are needed on pigment-protein complexes, membrane composition, and measurements of bacteria before firm conclusions can be drawn about the role of the NIR reflectance. Pigment absorbance features are strongly correlated with features of atmospheric spectral transmittance: P680 in PS II with the peak surface incident photon flux density at ~685 nm, just before an oxygen band at 687.5 nm; the NIR end of the red edge with water absorbance bands and the oxygen A-band at 761 nm; and bacteriochlorophyll reaction center wavelengths with local maxima in atmospheric and water transmittance spectra. Given the surface incident *photon* flux density spectrum and resonance transfer in light harvesting, we propose some rules with regard to where photosynthetic pigments will peak in absorbance: a) the wavelength of peak incident photon flux; b) the longest available wavelength for core antenna or reaction center pigments; and c) the shortest wavelengths within an atmospheric window for accessory pigments. That plants absorb less green light may not be an inefficient legacy of evolutionary history, but may actually satisfy the above criteria.








# DRAFT, Kiang, Astrobiology

## 1. Introduction

The utilization of the Sun's light energy by photosynthetic organisms provides the foundation for virtually all life on Earth, with the annual amount of carbon fixed from $CO_2$ into organic form by land plants and by ocean phytoplankton being each approximately ~45-60 Pg-C/yr (Cramer, et al., 1999, 2001), or 6-8% of the atmospheric carbon content (Reeburgh, 1997). The selective utilization of light energy results in two well-known spectral signatures exhibited in land plants: the "green bump," due to lower absorbance of green light by chlorophyll; and the "red edge," characterized by absorbance in the red by chlorophyll strongly contrasting with reflectance in the near-infrared due to refraction between leaf mesophyll cell walls and air spaces in the leaf. The red edge is so different spectrally from other matter that it has been measured by satellites to identify vegetation cover (Tucker, 1976; Grant, 1987; Sagan, et al., 1993) and estimate plant productivity (Potter, et al., 1993).

We seek to address and extend the age-old question: Why are plants green? More precisely, what is the functional role of different features of a photosynthetic organism's reflectance spectrum? Although it is fairly well understood how pigments absorb and how cells scatter light, it is not yet settled as to why photosynthetic pigments absorb at those particular wavelengths. Finally, how ubiquitous is the near-infrared (NIR) reflectance among photosynthetic organisms, and how does it serve the organism?

Photosynthetic spectral reflectance signatures are a result of both molecular constraints on biochemical processes and environmental pressures for adaptation. In this





DRAFT, Kiang, Astrobiology

review, we attempt to synthesize spectral characteristics across the full range of Earth's

photosynthesizers, covering the following:

1) a brief review of how photosynthesis works;

2) an overview of the diversity of photosynthetic organisms, their light harvesting

systems, and environmental ranges;

3) a synthesis of photosynthetic surface spectral signatures;

4) an exploration of evolutionary rationales for photosynthetic surface reflectance

spectra, including the Earth's chemical history, energy requirements for conversion of

photon energy to chemical energy, and the planetary light environment.

We conclude with some hypotheses about why photosynthetic pigments favor their

particular wavelengths, and whether alternative whole organism reflectance spectra could

be possible.  As this review is motivated by speculation about photosynthesis on Earth-

like planets in other solar systems, much of the discussion is placed in this context and

geared toward the diverse multi-disciplinary astrobiology audience. This review should

also be useful to specialists in Earth remote sensing, photosynthesis, and plant

physiology.

## 2.  Background:  Basic processes of photosynthesis, inputs and outputs

Photosynthesis efficiently converts light energy to electrochemical energy for

oxidation-reduction ("redox") reactions.  The excitation of light harvesting pigments by a





# DRAFT, Kiang, Astrobiology

photon of light causes an electron to be transferred along biochemical pathways that lead

to the reduction of $CO_2$. The electron is replaced by one extracted from the reductant.

The basic stoichiometry of photosynthesis is:

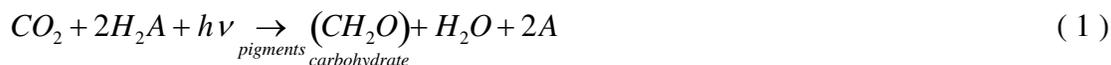

$$CO_2 + 2H_2A + h\nu \underset{pigments}{\rightarrow} (CH_2O)_{carbohydrate} + H_2O + 2A \qquad (1)$$

where $H_2A$ is a reducing substrate such as $H_2O$ or $H_2S$, and $h\nu$ is the energy per photon,

where h is Planck's constant, and $\nu$ is the frequency of the photon or the speed of light

divided by the photon wavelength.

When the reductant is water, then we have oxygenic photosynthesis:

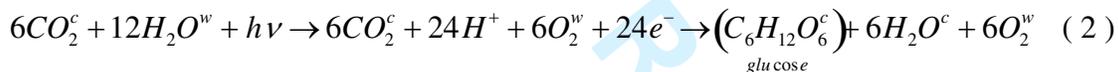

$$6CO_2^c + 12H_2O^w + h\nu \rightarrow 6CO_2^c + 24H^+ + 6O_2^w + 24e^- \rightarrow (C_6H_{12}O_6^c)_{glucose} + 6H_2O^c + 6O_2^w \quad (2)$$

where the superscripts c and w denote the oxygen from carbon dioxide versus that from

water. Four photons are required for each $O_2$ evolved (one photon for each bond in two

water molecules), and four photons are needed to reduce two molecules of the coenzyme

NADP+ eventually to reduce one $CO_2$. Thus, a minimum of 8 photons total are required

both to evolve one $O_2$ and to fix carbon from one $CO_2$. A few cycles are required to

obtain the six carbons to make the 6-carbon sugar, glucose. More than 8 photons are

generally required (experiments have shown up to about 12 photons; Govindjee, 1999)

because some are unsuccessful, and some in addition are used in the processes of cyclic

photophosphorylation (occurs in Photosystem I generation of ATP, described later; Joliot





# DRAFT, Kiang, Astrobiology

and Joliot, 2002; Munekage, et al., 2004) and nitrogen assimilation (Foyer and Noctor, 2002). The mechanisms by which these processes are achieved are highly complex, and the reader is referred to details in textbooks and recent findings (Voet, et al., 1999; Ke, 2001; Green and Parson, 2004; Ferreira, et al., 2004; Wydrzynski and Satoh, 2005).

Photosynthesis may use reductants other than water, such as $H_2S$, $H_2$, and $Fe^{2+}$ in anoxygenic photosynthesis. When the reductant is $H_2S$, then elemental sulfur is produced instead of oxygen, and that sulfur may be further oxidized to sulfate (Van Gemerden and Mas, 1995):

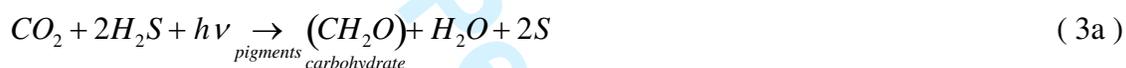

$$CO_2 + 2H_2S + h\nu \xrightarrow[pigments]{} (CH_2O)_{carbohydrate} + H_2O + 2S \qquad (3a)$$

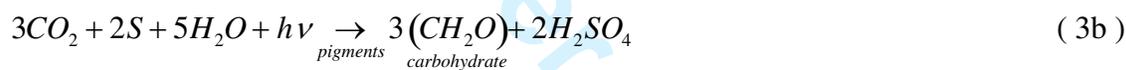

$$3CO_2 + 2S + 5H_2O + h\nu \xrightarrow[pigments]{} 3(CH_2O)_{carbohydrate} + 2H_2SO_4 \qquad (3b)$$

When the reductant is $H_2$ (Vignais, et al., 1985) or $Fe^{2+}$ (Jiao, et al., 2005; Ehrenreich and Widdel, 1994), the reactions are:

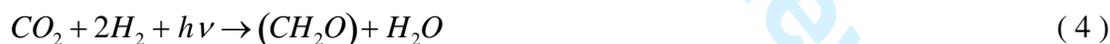

$$CO_2 + 2H_2 + h\nu \rightarrow (CH_2O) + H_2O \qquad (4)$$

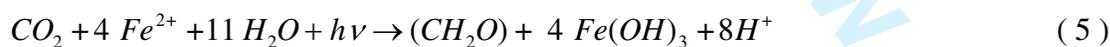

$$CO_2 + 4\ Fe^{2+} + 11H_2O + h\nu \rightarrow (CH_2O) + 4\ Fe(OH)_3 + 8H^+ \qquad (5)$$

(In Equation 5, actually $HCO_3^-$ and not $CO_2$ is directly used, though it could come from $CO_2$ dissolved in water or from dissolution of some other carbonate source; the $Fe^{2+}$ could be from dissolution of, e.g. $FeCO_3$).

The quantum requirement in all these cases is also 8-12 photons per carbon fixed. In summary, the inputs to photosynthesis are light energy, a carbon source, and a







DRAFT, Kiang, Astrobiology

reductant. The direct products are carbohydrates and can be oxygen, elemental sulfur,

water, and other oxidized forms of the reductant.  Soil nutrients are also necessary inputs,

described in more detail below.

## 3.  Range of photosynthetic organisms, habitats, pigments, metabolisms, and environmental limits

Photosynthetic organisms are adapted to occupy different niches according to

both physical (light, temperature, moisture) and chemical resource (electron donors,

carbon sources, nutrients) requirements.  In addition, their environmental ranges may be

controlled by the presence of competing or grazing organisms.  The range of

photosynthetic organisms is shown in Table 1, which lists the approximate time of their

appearance on Earth, the significant features that distinguish them metabolically and

spectrally, their relative abundance and productivity over the Earth, habitats, their

pigments types, reaction center types, electron donors, growth mode and growth form,

carbon sources, and metabolic products.

**Emergence of photosynthetic organisms on Earth.**  Photosynthesis arose fairly

early in the history of the Earth.  One theory about the origin of photosynthesis is that it

began as a fortuitous adaptation of primitive pigments for infrared thermotaxis of

chemolithotrophic bacteria in hydrothermal ocean vents (early Archean, possibly 3.8 Ga

ago; Nisbet, et al., 1995).  Thus, less dependent on the heat of the hydrothermal vents, the





DRAFT, Kiang, Astrobiology

habitats of these organisms gradually expanded to shallower waters where solar light could be utilized (Nisbet and Fowler, 1999; Des Marais, 2000). It is also possible that photosynthesis arose first in shallow waters and, before there was oxygenic photosynthesis to provide an ozone shield, UV screening proteins led to the transfer of excitation energy to the porphyrin (Mulkidjanian and Junge, 1997). The evolution of ocean chemistry could have played a major role in the evolution of chlorophyll (Mauzerall, 1976; Dasgupta, et al., 2004; Dismukes, 2001; Blankenship and Hartman, 1998). Blankenship and Hartman (1998) proposed that hydrogen peroxide, $H_2O_2$, could have been a transitional electron donor on the oxygen-poor early Earth. Liang, et al. (2006) posited that the necessary high $H_2O_2$ in the oceans could have occurred after a "Snowball Earth" event (low-latitude glaciation), around 2.3 Ga, due to storage of $H_2O_2$ in ice and its subsequent release into the oceans upon de-glaciation. Protocyanobacteria early in the Archean may have utilized $Fe(OH)^+$ as a reductant (Olson, 2006). Dismukes, et al. (2001) proposed that bicarbonate could have been a transitional reductant before water in the Archean ocean and reacted with manganese (II) to form Mn-bicarbonate clusters as precursors to the $(Mn)_4$ core of the oxygen evolving complex; the redox potentials are such that green sulfur bacteria would have been the early hosts of these clusters prior to the evolution of cyanobacteria.

The fossil record seems to support that the early photosynthesizers were anoxygenic purple bacteria and green sulfur bacteria that used reducing substrates other than water, such as $H_2$ and $H_2S$ (Olson, 2006). Anoxic environments allowed these organisms to thrive, since oxygen is damaging to bacteriochlorophylls. By the mid- to late-Archean (3.5-3.6 Ga ago) or as late as 2.3 Ga ago in the Early Proterozoic, oxygenic





DRAFT, Kiang, Astrobiology

cyanobacterial mats had formed in the shallower waters. Endosymbiosis of cyanobacteria and early animal protists gave rise to the plastids of algae, these plastids being the photosynthesizing organelle (Larkum, et al., 2003). The algae were, thus, the first eukaryotic photosynthesizers and occur in both single-cell and multi-cellular forms (e.g. kelp). Red algae appeared around 1.2 Ga ago. Eukaryotic green algae did not appear until as late as 750 Myr ago. These photosynthetic organisms were protected under water from UV radiation until $O_2$ and, hence, $O_3$ built up in the atmosphere. With the rise of atmospheric $O_2$, the anoxygenic ancestors lost their competitive advantage to organisms that could withstand and respire $O_2$ (or so one may interpret). Land plants are believed to be descended from a single branch of the green algae (Larkum, et al., 2003). The first land plants, with the non-vascular Bryophyta (mosses and liverworts) being the earliest, did not occur until 460 Myr ago, after which the abundance of plant life exploded, reaching a peak of productivity during the Carboniferous period 354 Myr ago (Carroll, 2001). Further complexity continued to arise through the emergence of flowering plants (144 Myr ago), and more water-efficient photosynthetic pathways came about through anatomical and enzymatic changes. In crassulacean acid metabolism (CAM) photosynthesis, which arose 70-55 Myr ago, $CO_2$ is stored in an intermediate at night; during the day, stomates remain closed to prevent water loss, and the intermediate is broken down for $CO_2$ to enter the Calvin cycle. In C4 photosynthesis, which arose 20-35 Myr ago, $CO_2$ is similarly concentrated in an intermediate that is transported to an internal bundle of cells that allow stomates to draw in atmospheric $CO_2$ through a smaller aperture (Sage, 2001).





DRAFT, Kiang, Astrobiology

So it seems natural for organisms to evolve to capture stellar energy, and Earth's example shows that, they continue to get better at it, especially once they emerge from the water. Earlier organisms are more diverse in their photosystems, while later organisms are more complex in their morphological properties. Note that not all photosynthesizers are autotrophs (fix $CO_2$), but many of the bacteria are heterotrophs that utilize organic carbon, though some may use both inorganic and organic carbon. The halobacteria do not perform actual photosynthesis, in that no electron transfer is performed, but their pigment bacteriorhodopsin drives a proton pump for heterotrophic assimilation of organic carbon. Of greatest interest to us are those photosynthesizers that are autotrophic and can fix $CO_2$ directly, as these are the primary producers and the foundation for life on Earth. We survey next their photosystems and environmental constraints.

**Light harvesting.** Photosynthetic organisms have intricate photosystems that coordinate 1) the spectral selection of light energy and 2) the abstraction of electrons from electron donors. These photosystems are composed of three main components. The 1) *peripheral* or *outer antenna* complex transfers light energy to 2) the *core* or *inner antenna*. These two together are known as the light-harvesting complex (LHC). The core antenna is also an integral part of the third component, 3) *reaction center* (RC) complex, where light energy is finally converted to chemical energy in charge separation (e.g. with $H_2O$, other electron donors, enzymes) (Ke, 2001).

The photon utilized must be of sufficient energy to generate a voltage potential difference that is great enough to oxidize the reductant as well as afford the electron





DRAFT, Kiang, Astrobiology

transfers for reduction of the relevant intermediates (and some enzymes). In plants and all oxygenic photosynthesis, there are two stages of light utilization: 1) in the extraction of electrons from water (using Photosystem II or PS II, peaking in absorbance at 680 nm) to regenerate the next step; and 2) for the reduction of the electron carrier $NADP^+$ (using Photosystem I or PS I, peaking in absorbance at 700 nm), which is then used in the Calvin-Benson Cycle and for the synthesis of ATP. This two-step sequence is known as the "Z-scheme" for the zig-zag in redox potential at each step (first proposed by Hill and Bendall, 1960; first proven by Duysens, et al., 1961; Blankenship and Prince, 1985). The redox potential is the Gibbs free energy change of a reaction (calculated with the Nernst equation), which may be expressed in volts and is the propensity of an oxidation-reduction reaction to proceed spontaneously in one direction or the reverse; the convention is that hydrogen has a redox potential of zero (at equilibrium conditions) and electrons move spontaneously in the direction of higher (toward positive) potentials. These thermodynamics determine how much energy may be stored as product, while the kinetics of electron transfer affect light harvesting and the quantum yield. All oxygenic photosynthesizers utilize both PS I and PS II, with water as the reductant. Anoxygenic photosynthesizers utilize only a single photosystem that uses other electron donors, for which equations were given earlier. Figure 1 shows the reaction center midpoint redox potentials (when concentrations of reductants and oxidants are the same) and excited potentials for the Z-scheme of oxygenic photosynthesis, as well as the electron transport pathways for different classes of bacteria. The Type I and Type II reaction center categories distinguish the types of electron acceptors used. Only oxygenic





DRAFT, Kiang, Astrobiology

photosynthesis is known to utilize two reaction centers in sequence. For a more detailed

overview, the reader is referred to Blankenship (2002).

Photosynthetic pigments are categorized into three chemical groups or

"chromophores": the chlorophylls, carotenoids, and phycobilins. The chlorophylls are

the pigments of the reaction centers and also occur in the core antennae and light

harvesting antennae. Specialized chlorophylls at the RCs serve to "trap" the excitation

energy and convert the electronic energy to chemical energy through charge separation.

The funneling of energy from the LHCs in all eukaryotes is achieved through a rather

remarkable process known as *resonance excitation transfer* (also known as the Förster

mechanism; review: Clegg, 2004), in which a pigment is excited by light at a particular

wavelength and the subsequent de-excitation of the pigment, rather than resulting in a

loss of energy to heat or fluorescence, leads to the excitation of another pigment whose

energy level overlaps. A series of such excitations and de-excitations creates an "energy

cascade" toward longer wavelengths. In all oxygenic eukaryotes, Chl a occurs in the core

antenna and acts as the primary donor in the reaction center; $H_2O$ is the electron donor.

Other chlorophylls (Chl b/c/d) provide light harvesting roles, but very recently Chl d,

which was discovered in cyanobacteria (Miyashita, et al., 1996; Miller, et al., 2005), may

replace Chl a in the reaction centers in some cyanobacteria that live in environments with

little visible light (Larkum and Kühl, 2005; Chen, et al., 2005;). Chl d has its major peak

absorbance in the NIR at ~720 nm (Manning and Strain, 1943; Larkum and Kühl, 2005),

and thus, oxygenic photosynthesis is being performed in the NIR! In non-oxygenic

bacteria, bacteriochlorophylls play the role of primary donor, with a great variety having





DRAFT, Kiang, Astrobiology

distinct absorption spectra; a variety of electron donors are possible, including $H_2S$, $H_2$,

Fe, sulfate, thiosulfate, sulfite, and organic carbon.

The other pigments as well as additonal chlorophylls (particularly Chl b in plants

and green algae, and Chl c in diatoms and brown algae) in the light harvesting complexes

(LHCs) act as "accessory" pigments to help obtain light energy, the carotenoids in the

blue and green, and the phycobilins (cyanobacteria and red algae) in the green and

yellow. The carotenoids in addition serve to protect chlorophyll against photo-oxidative

damage in conditions of high light, high temperature, $O_2$, and the presence of certain

pigments (Nobel, 1999). Oxygen would be toxic to photosynthetic organisms without the

presence of carotenoids.

The absorbance spectra of all pigments are influenced by the protein complexes in

which they are bound, such that there can be variations in the same type of pigment and

the peak absorbances *in vivo* tend to be broadened and shifted compared to those of the

pure pigments extracted in solution (Vasil'ev and Bruce, 2006). For example, Chl a in

PS I peaks at 700 nm, whereas in PS II it peaks at 680 nm. We compare *in vivo* pigment

spectra (in intact membranes) below.

Figure 2a shows the *in vivo* absorbance spectra of the major pigments found in

plants and algae (Chl a, Chl b, carotenoids; phycobilins exist only in certain algae and

cyanobacteria), as well as the spectrum of Chl a fluorescence (Papageorgiou and

Govindjee, 2004), together with the incident solar radiation spectrum at the top of the

Earth's atmosphere and at the surface of the Earth after atmospheric filtering. Sources

and measurement methods for the pigment spectra are summarized in Appendix A1. The

Chl a fluorescence spectrum effectively identifies the upper wavelength limit to





DRAFT, Kiang, Astrobiology

absorbance by chlorophyll. The photon flux densities are from models, as detailed in the

figure caption; in addition, measurements from buoys at Hawaii (Dennis Clark, NOAA)

are plotted to show how varying atmospheric optical depth can affect the incident light

spectrum. The atmosphere performs spectral filtering of radiation, with Mie and

Rayleigh scattering of light in the bluer wavelengths, and with clear bands of absorption

by gases, most importantly $O_3$, $O_2$, $H_2O$ vapor, and $CO_2$, as indicated in the figure. Plant

and algae pigments clearly must have evolved to have their absorbance peaks in

atmospheric transmittance windows for radiation and are confined to the visible range,

which is commonly considered to be "photosynthetically active radiation" (PAR), 400-

700 nm (the tail of the absorbed range extends as far as 730 nm).

Figure 2b shows the *in vivo* absorbance spectra of the bacteriochlorophylls with

associated carotenoids in intact membranes (sources and measurement method

summarized in Appendix A1). Above these are plotted the transmitted radiation

spectrum through 5 cm of pure water as well as through 10 cm of water containing algae

(water absorption coefficient from Segelstein, 1981, Sogandares and Fry, 1997, and Kou,

et al., 1993; algae absorption coefficient from kelp, see Appendix A3 for explanation of

calculation). Water is highly transmitting in the visible and highly absorbing in the NIR,

as can be seen from the spectrum for pure water at 5 cm depth. A small transmittance

window exists, however, in the NIR, peaking at 1073 nm. Algae and cyanobacteria in the

upper layers of water may strongly attenuate visible light, such that only radiation above

about 700 nm may be transmitted to depth and attenuated again by water above 900 nm.

The cyanobacterium, *Acaryochloris marina*, which uses Chl d at ~715-720 nm instead of

Chl a, may be adapted to receive this remaining longer wavelength filtered through





DRAFT, Kiang, Astrobiology

overlying organisms.  The bacteriochlorophylls commonly absorb in the range 700-900

nm.  In an extreme case, BChl b in the purple bacteria *Blastochloris viridis* (formerly

*Rhodopseudomonas viridis*) can absorb to wavelengths as long as 1013-1025 nm (Scheer,

2003;  Trissl, 1993).  *B. viridis* inhabits murky, anoxic sediments where little visible light

penetrates.  Its exact peak absorbance is sensitive to temperature, and the range of

measurements on lab cultures may, perhaps, not be exactly the same as in the bacterium's

native environment.  The local peak in water transmittance at 1073 nm leaves room for

speculation that either *B. viridis* has the capability to harvest light at even longer

wavelengths, or the wavelength of peak absorbance is limited by exciton transfer kinetics

to the reaction center or thermodynamic constraints.

In addition to the bacteriochlorophylls, phototrophic bacteria will also utilize

carotenoids, with the ratio between bacteriochlorophyll and carotenoids varying with

light quality.  In Figure 2b, the absorbance peaks in the blue wavelength range are due to

carotenoids.  Purple and green sulfur bacteria pigments thus absorb in transmittance

windows under water.  Overall, the full range of pigments has been observed to absorb

light in the wavelength ranges 330-900 nm and 1000-1100 nm (Scheer, 2003). Oxygenic

photosynthesis on Earth is limited to photosystems that operate at 400-730 nm, but

anoxygenic photosynthesis occurs at wavelengths as long as 1015-1020 nm (Trissl, 1993;

Scheer, et al, 2003).

To summarize (see Table 1), phototrophic bacteria utilize bacteriochlorophylls in

their reaction centers and are the most diverse in their growth modes, many being

heterotrophs and some using sulfide, iron, or hydrogen as their electron donors (Eraso

and Kaplan, 2001; Ehrenreich and Widdel, 1994).  Among the phototrophic bacteria, only





DRAFT, Kiang, Astrobiology

cyanobacteria (formerly known as blue-green algae) are oxygenic.  The diversity of

pigments of cyanobacteria includes Chl a/b/c/d, phycobilins, and carotenoids, which

allows them a wide range of colors and distribution nearly everywhere on Earth, from

aquatic environments to desert salt crusts.  Where cyanobacteria or algae coexist with

fungi in lichens, they can support a level of productivity comparable to vascular plants.

All algae utilize Chl a and carotenoids; in addition, green algae use Chl b.  Several other

algae (e.g diatoms and brown algae) use different forms of Chl c, and red algae utilize

phycobilins instead of Chl b or Chl c.  Plants utilize only Chl a, Chl b, and carotenoids

but have developed more complex mechanisms to acquire $CO_2$ and retain water.  All

photosynthetic organisms have carotenoids.

Because the photosystems are tied to the type of electron donor, the

photosynthesizer is therefore necessarily restricted to particular resource environment

niches. Purple and green sulfur bacteria are restricted to aquatic environments where

sulfur (sulfide, sulfate, sulfite, thiosulfate, $H_2S$) is available as an electron donor,  and

some of these bacteria require anoxic environments.  Cyanobacteria and other

photosynthetic bacteria are also known to form mutualistic communities in microbial

mats in benthic (Decker, et al., 2005) and freshwater aquatic environments (Wiggli, et al.,

1999), where distinct layering demarcates both light and chemical niches.  Exudates and

organic carbon from one layer provide the electron source or food source for another, and

the lower, anoxic layers utilize sulfide or $H_2S$ for the electron donor.  The upper layers of

cyanobacteria and algae absorb visible light, such that the lower layers of sulfur bacteria

utilize the NIR radiation that transmits through.





DRAFT, Kiang, Astrobiology

Next we will comment briefly on the other environmental constraints on photosynthetic organisms, as these can affect also their spectral properties.

**Climate – water and temperature.**  Water and temperature are the primary constraints on the distribution, spatially and temporally, of different organisms.  For plants, the reader is referred to well-known global surveys of the climate limits of, for example, grasslands, broadleaf temperate deciduous forests, evergreen needleleaf forests, boreal rainforest, to name some possible categories of many (Holdridge, 1967; Whittaker, 1975;  Woodward, 1987;  Larcher, 1995).  These large-scale classifications show clear correlations between climate and plant form and mixtures of plant communities, such as broadleaf trees in temperate and tropical zones versus needleleaf trees in colder climates.  The resulting difference in spectral signatures can be distinguished by satellites (Tucker, et al., 1985; Defries and Townshend, 1994).  Mosses and liverworts, which lack vascular structure, require very moist and often shaded environments, and do not form large structures.  Lichens (symbioses between fungi and algae or cyanobacteria) are limited to environments with moist air but can be highly productive. They are often the first colonizers on rock substrate and can be the dominant source of net primary productivity and a source of food for animals in some land ecosystems (Ager, et al, 1987; Rees, et al., 2004), in some cases accounting for 70% of the land cover (Solheim, et al., 2000).  The extreme temperature limits of photosynthetic organisms range as low as -15.7 ℃, the survival limit for  arctic snow algae and arctic ice shelf cyanobacteria (but they require liquid water for growth) (Gorton, et al., 2001; Mueller, et al., 2005) and as high as ~75℃ for  bacteria in hot springs (Miller, et al., 1998).  Meanwhile, the temperature limits for Earth life in general can be as low as -20





# DRAFT, Kiang, Astrobiology

°C (possibly as low as -196 °C for some methanogenic bacteria) (Junge, et al., 2006) and

possibly as high as 120 °C (Kashefi and Lovley, 2003).

**Extreme environments.** Photosynthetic bacteria and some algae tend to inhabit

what on Earth are considered extreme or stressful conditions, such as low light (e.g. *B.*

*viridis*), desert (Karnieli, et al., 1999), snow and ice (algae, Gorton, et al., 2001;

cyanobacteria, Mueller, et al., 2005), saline (Decker, et al., 2005), low pH (pH of 0, red

algae, Schleper, et al., 1995; archaea at pH 0, Edwards, et al., 2000), high pH (pH of 10,

cyanobacteria, Finlay, et al, 1987), and anoxic environments.

**Light quantity.** Some organisms may function better in low or high light

environments. The lower light limit is determined by the balance between photosynthesis

and respiration, the cut-off between survival and death (the "light compensation point"),

while the upper limit is determined by other resource limits (water, nutrients, carbon

source) and the ability to protect against photo-oxidation (damage to chlorophyll due to

excited states of $O_2$ in high light). Rather than quantifying light in terms of energy, we

express it here in photons (or moles of photons) because photosynthesis depends on

photon flux (at particular wavelengths) rather than energy flux. The lowest observed

light compensation points are ~ 3 μmol-photons $m^{-2}$ $s^{-1}$ (0.7 $W/m^2$, 1.8 x $10^{18}$ photons $m^{-2}$

$s^{-1}$) of PAR for green plants (Nobel, 1999) and ~ 0.01 μmol-photons $m^{-2}$ $s^{-1}$ red macro-

algae (6 x $10^{15}$ photons $m^{-2}$ $s^{-1}$, Littler, et al., 1986). Overmann, *et al.* (1992) observed a

brown sulfur bacterium, living at ~80 m depth in the Black Sea, that is adapted to an

available irradiance of 0.003-0.01 μmol-photons $m^{-2}$ $s^{-1}$ (1.8-6.0 x $10^{15}$ photons $m^{-2}$ $s^{-1}$)

(additional characterization by Manske, et al., 2005). The theoretical unicellular light

limit has been estimated by Raven (1984) to be ~ 0.1 μmol-photons $m^{-2}$ $s^{-1}$ (6 x $10^{16}$





# DRAFT, Kiang, Astrobiology

photons $m^{-2}$ $s^{-1}$), and as this is higher than that observed, additional efficiency strategies

for survival at low light must be more a possibility than current understanding allows.

Recently, Beatty, *et al.* (2005) discovered a green sulfur bacterium that utilizes

geothermal light at a hydrothermal vent, where the irradiance close to the vent is

comparable to that at the 80 m depth of the Black Sea. This opens up the possibilities for

photosynthesis independent of starlight. However, it is unlikely that these bacteria

evolved under such low light conditions, but we think they probably are migrants from

surface waters. For the upper limit of photon flux density, Wolstencroft and Raven

(2002) summarized the literature and found a theoretical tolerance for land plants against

photodamage at 6-9 mmol-photon $m^{-2}$ $s^{-1}$ (3.6-5.4 x$10^{21}$ photons $m^{-2}$ $s^{-1}$) over the PAR

band, which is well above Earth's typical flux of 2 mmol photon $m^{-2}$ $s^{-1}$ (1.2 x $10^{21}$

photons $m^{-2}$ $s^{-1}$). For Earth-like planets in general, they conjectured a theoretical upper

limit for land organisms to be 10 mmol photon m-2 s-1 (6 x 1021 photons m-2 s-1).

Since aquatic organisms are shielded under water, they could exist for even higher

surface photon flux densities.

**Ultraviolet light damage.** UV light is damaging to DNA, and exposure to UV at

levels received at the Earth's surface generally inhibits photosynthesis and leaf expansion

(Karentz, et al., 1994; Nobel, 1999; Tenini, 2004). Therefore, most of the discussion on

UV effects on habitability focuses on protection against UV by ozone or

microenvironments (Cockell and Raven, 2004; Cockell, 1999; Kasting, et al., 1997;

behaviors such as photomotility to avoid the harmful UV (Bebout and Garcia-Pichel,

1995); and specialized screening pigments or phenolics that prevent the penetration of

UV light into the cell (scytonemin in arctic cyanobacteria, Mueller, et al., 2005;





# DRAFT, Kiang, Astrobiology

phenolics in snow algae, Gorton, et al., 2001; mycosporine-like amino acids in response

to UVB photoreceptors in cyanobacteria, Portwich and Garcia-Pichel, 2000).  Such

protective compounds may have played a role in the early emergence of plants onto land

(Cooper-Driver, 2001; scytonemin found in ancient rocks by Marilyn Fogel, personal

communication, NASA Astrobiology Institute conference, 2005).  The maximum flux of

UV at the Earth's surface is $1.8\text{-}2.8\text{x}10^{18}$ photons/m$^2$/s over the UV-B (280-315 nm) band

at the Equator at noon under cloudless conditions, or averages globally 0-12 kJ/m$^2$/day

($2.1\text{x}10^{17}$ photons/m$^2$/s, converting with the average energy per photon in the UV-B).

Damage to plants from UV-B radiation has been observed at doses of 15-16 kJ/m$^2$/day

($2.6\text{-}2.8\text{x}10^{17}$ photons/m$^2$/s, Kakani, et al., 2003).

   **Nutrients.**  Nutrients, as mentioned earlier, are also limiting resources.  The

nature of the limitation is on level of productivity and competitive advantage, rather than

physiological tolerance.  On land, the succession of species is often a function of the

long-term development of the soil, which progresses through the release, addition, and

eventual occlusion or loss of minerals and nitrogen. A comprehensive review of nutrient

cycling is beyond the scope of this paper, but we summarize the most important limits

here. Fixed N (available in the soil as $NO_3^-$ and $NH_4^+$)  and minerals P, K, S, Mg, Fe, Mn

are the nutrients required for the production of pigments and enzymes, with N and P

generally being the most limiting nutrients.  Nitrogen is the main limiting nutrient, and its

content in photosynthesizers is the prime correlating variable with photosynthetic

capacity (Schulze, et al. 1994), since the chlorophylls are tetrapyrroles with four nitrogen

atoms surrounding a magnesium atom (Nobel, 1999).  Nitrogen must be fixed originally

from atmospheric $N_2$ by enzymatic processes that occur in only particular organisms and





DRAFT, Kiang, Astrobiology

form $NO_3^-$ and $NH_4^+$. Availability is constrained, on land, by ecosystem age (time

required for nitrogen fixing organisms to input nitrogen from the atmosphere into

developing soil) and losses by leaching or fire. In aquatic and marine environments,

availability is constrained by diffusion from the atmosphere and deposition from land-

surface run-off of organic compounds. Mineral nutrient availability is locally constrained

by rock substrate, geothermal sources, and deposition from non-local sources.

Phosphorus, an essential mineral for DNA, ATP, and phospholipids of cell membranes,

becomes available from weathering of the mineral apatite but then, over time, complexes

with Al, Fe, and Mn (at low pH) or with Ca (at high pH), such that it becomes

unavailable. In aquatic or ocean environments, Fe and P in addition to N are the primary

limiting nutrients, and are input via atmospheric deposition or river run-off. For more

details, the reader is referred to Schlesinger (1997).

## 4. Spectral signatures of photosynthetic organisms

Given the above information on photosynthesizers' metabolism, pigments, and

environmental niches, how do these features combine into the spectral reflectance of the

whole organism? Figure 2c shows the surface incident photon fluxes of the Sun/Earth (in

millimoles) with reflectance spectra of an oak leaf, a grass, a moss, and a lichen. On first

glance, the red edge features appear to show some striking correlations with atmospheric

absorbance features, particularly oxygen and water bands. We wish to find a

physiological explanation. Physical explanations of land plant spectral signatures are





DRAFT, Kiang, Astrobiology

fairly well understood in some aspects, whereas there is less of such information on other
photosynthesizers.   In this section, we first detail the reflectance properties of plant
leaves.  We then compare the reflectance spectra of other photosynthetic organisms to
identify evolutionary pressures that would lead to variations, if any, particularly with
regard to the "red edge."

### 4.1. Components of plant leaf spectral reflectance, physical explanation

Grant (1987) and Vogelmann (1993) review in detail the optical properties of
plant leaves, so only a brief summary from these reviews is provided here.  In addition,
the reader is referred to Tucker (1976, 1978) for discussion of satellite sensor bands for
monitoring whole vegetation canopies.

Figure 2c shows the typical reflectance signature of land plants, for which the
significant features include the green bump in reflectance and the "red edge."  The latter
is so-called because plant photosynthetic pigments absorb strongly in the visible or PAR,
which strongly contrasts with high scattering in the NIR due to refraction between leaf
mesophyll cell walls and air spaces inside the leaf.  The wavelength of the "red edge" is
more strictly defined as the inflection point in the slope of the reflectance between the red
and NIR, and is sometimes referred to as the "red edge inflection point" or the "red edge
position."  It falls generally around 700 nm, but the location and steepness may vary
according to the organism's abundance or thickness (if sensing over a large area) and
physiological status; as shown later in this paper, there can be distinct differences
between organism types.





DRAFT, Kiang, Astrobiology

Using the leaf radiative transfer model of Jacquemoud and Baret (1990), we

illustrate in Figure 3  how a leaf's spectral reflectance can vary according to a) leaf

structure and content of b) water, c) Chl a and b, and c) carbon.  The leaf structure is the

main determinant of the high reflectance in the NIR in the overall signature, as leaf

thickness and interstitial air spaces between mesophyll cells determine the surfaces off of

which NIR is scattered. The cell membranes are composed of lipids and proteins, and the

cell walls of cellulose and lignin. The refractive index in crop plants has been measured

as 1.333-1.48, with averages around 1.43 (refractive index of air is 1.0, water 1.333,

soybean oil 1.48) (Gausman, 1974;  wavelength-dependent refractive index: Jacquemoud

and Baret, 1990).  Note that, in fact, the cell walls also scatter visible light (Figure 3c,

gray line with no Chl a & b), but pigments absorb here when present.  Water content

affects reflectance in the longer wavelengths, with strong absorbance bands at ~1400 nm

and ~1900 nm.  In fact, the exact locations of these bands in the organism may shift with

physiological status.  Hydration status of a leaf positively affects the NIR reflectance,

since cell turgor affects air spaces within the leaf.  The Chl a and b content, of course,

affects the absorbance in the visible.  The carbon density affects just the reflectance

bands in the NIR.  The absolute red/NIR contrast, in general, increases for thicker leaves

or multiple layers of leaves or organisms, as more chlorophyll per area will absorb more

in the visible, and more layers and interstitial air spaces increase the surfaces for NIR

scattering per unit area.  A weaker observed contrast with the NIR may be possibly due to

properties of the organism surface, which we will detail later when comparing organisms.

The model of Jacquemoud and Baret (1990), based on a plate representation of

the leaf (leaf structure is just a tunable parameter), represents well the vertical variations





DRAFT, Kiang, Astrobiology

in the spectral reflectance in response to the given parameters.  It does not deal with horizontal variations (wavelengths of critical features) such as:  variation in other pigments that affect the visible spectrum; observations that the NIR end of the red edge may broaden in curvature and shift in location; and observations that the water absorbance bands may blue- or red-shift with physiological status due to, for example, water stress or phenological changes (Filella and Penuelas, 1994;  Penuelas and Filella, 1998; Karnieli, et al., 1999). The nature of these shifts in the NIR reflectance is not well studied.

## 4.2. Variations in reflectance spectra among photosynthetic organism types

Now, we offer perhaps the first attempt at a comprehensive comparison of reflectance spectra across photosynthetic taxa to determine whether this will yield further insights into how such spectra may have evolved.  In particular, we focus on how the red edge varies among organisms, because this is the strongest feature that spectrally differentiates photosynthesizers from the background surface or water.  The red edge begins at about 680 nm, where chlorophyll a peaks in its absorbance in Photosystem II (PS II) (antenna of both PS II and PS I absorb almost equally 680 nm light) and then the reflectance plateaus around 720-760 nm.  An evolutionary explanation for why these features occur at those particular wavelengths is not settled, particularly with regard to the NIR reflectance.  Our driving questions here are: where does the red edge begin, where does it end, and where could it occur on another Earth-like planet, if at all?





DRAFT, Kiang, Astrobiology

Since the red edge is not merely a step function but has a slope with a bottom in the visible and a top in the NIR, various workers have tried to quantify the location of the red edge to distinguish species or quantify variations due to physiological status. The first derivative of the reflectance with respect to wavelength is a common measure by which to identify the point of maximum slope of the red edge, the "red edge inflection point" (REIP). Some workers take this point as the strict definition of the red edge, but here we will use "red edge" to encompass the span of the rise in reflectance from the visible to the NIR. The REIP varies with the level of the red absorbance and NIR reflectance, as well as with shifts in the wavelength of the onset of the NIR plateau. Therefore, we hypothesize that the wavelength of the NIR end may have more physical meaning than the red edge inflection point. For this wavelength of onset of the NIR plateau, we will coin the term, the "NIR end."

This NIR end is at least a function of the spectral spread of pigment absorbance, which, as we mentioned earlier, is affected by the proteins in which the pigments are bound. We are interested in surveying the variation in this NIR end and discerning whether it is the result of any environmental adaptations. Quantification of the location of this NIR end is not entirely straightforward because of the variations in curvature and slope that can occur over the NIR end and NIR plateau, respectively. We found the third derivative of the reflectance with respect to wavelength to capture fairly well the point at which the NIR plateau begins to level off, as this quantifies the "jerk" or change in the acceleration of the slope. The second derivative identifies better the point just before the NIR onset begins, rather than the plateau side. Because these measures are still





DRAFT, Kiang, Astrobiology

somewhat subjective, we also calculate the first derivative as a well-defined measure of at least relative differences between spectra.

**Presence of the red edge across photosynthetic taxa.** Figure 4 shows spectral reflectance measurements (not a model) of a) land-based vascular plants, b) aquatic plants, c) mosses, d) lichens, e) algae, and f) different layers of a microbial mat in an alpine lake (species and sources summarized in Appendix A2). From this survey, it appears that all photosynthesizers except the purple bacteria (Figure 3d) have a "red edge," albeit this edge is weak in lichens and bacteria. For organisms under water, the NIR plateau in the range 760-850 nm would be absorbed by water and not visible from above, but apparently, as seen from the aquatic plants' spectra here, when removed from the water, the refractive properties with the air are such that these organisms also have a red edge.

In stark contrast, Figure 4e shows that the purple bacteria clearly have no red edge, but instead show strong absorbance in the NIR and possibly adjacent NIR "edges." The green sulfur exhibit red edge-like reflectance, with variation in the pigments in the visible. The orange line is of a mixture of filamentous bacteria, diatoms, and precipitations of elemental sulfur; a red-edge feature also appears in this mix. All spectra in this figure are *in situ* measurements of whole mat layers (but outside the water) of species from the same microbial mat in an alpine bog pond (Wiggli, et al., 1999; data courtesy of Reinhard Bachofen). The purple bacteria are *Chromatium* species, anoxygenic photolithoautotrophs with BChl a or BChl b, and engage in sulfide reduction. The green sulfur bacteria are also anoxygenic and utilize BChl a and BChl c, d and e, the latter three bacteriochlorophylls having absorption peaks in the range 718-750 nm. The





DRAFT, Kiang, Astrobiology

colonies may be mixed with oxygenic bacteria, such that the spectrum is not purely green

sulfur bacteria. More measurements are needed to confirm distinct spectra for these

different kinds of bacteria. Figure 4g shows reflectance spectra of various non-

photosynthetic surfaces, mineral and organic, including human skin and dead grass.

Hematite and (live) human skin show some striking similarities to the photosynthetic

organism spectra in that they have a color in the visible and high reflectance at longer

wavelengths, but which, are very different from those of the red edge features.

**Horizontal variations in the NIR end across taxa.** Figure 5 shows enlargements

of the red edge section of the reflectances in Figure 4, with the 761 nm oxygen absorption

line as a reference that reveals differences in the wavelength of the NIR end. As an

example of how we calculate the NIR end, Figure 6 shows the third derivative of

reflectance (left plot) and the reflectance spectrum of an aquatic plant. The vertical line

indicates the maximum of the third derivative and the NIR end. Figure 7 shows a scatter

plot of a) the NIR end wavelengths of the spectra by organism type, and b) the maximum

slope wavelengths for the same spectra. There are clear trends by organism type: the

terrestrial plant NIR end is reddest, ranging 746-765 nm; aquatic plants cluster blue-ward

of land plants at 730-745 nm; mosses, temperate lichens, and temperate algae range over

720-733 nm. The snow algae have the bluest red edge. Because of the noisiness of the

snow algae data, we could not calculate the NIR end, but the maximum slope wavelength

shows clear trends. Also, due to interference of atmospheric oxygen, some reflectance

spectra have a spike at 761 nm.

From this survey, it appears that the NIR reflectance, though common, does not

appear to be universally the same among all photosynthetic organisms, but there appear





DRAFT, Kiang, Astrobiology

to be consistent variations among taxa. Lichens often have no sharp edge, but a steady slope. Purple bacteria have possibly an NIR edge, which is consistent with their pigment absorbance spectra. There seems to be a trend in the "NIR end" (where the NIR reflectance begins to plateau in the red edge), in which the most structurally advanced to simplest organisms are ordered from reddest to bluest. More research is needed to explain these trends.

## 5. Evolutionary Rationales for Photosynthetic Surface Spectral Reflectance

How the properties of pigments, cell membranes, and cell walls evolved is not well known, but some evolutionary rationale is needed, with regard to why particular spectral features of photosynthetic organisms occur at particular wavelengths, before we can conjecture where similar features might arise on another planet. From what we have observed of Earth photosynthesizers, it seems especially important to know the evolutionary pressures on the following features: 1) photosynthetic reaction center excitation wavelengths, 2) core antenna peak absorbances, 3) accessory pigment peak absorbances, 4) beginning and ending wavelengths of the red edge, 5) NIR reflectance bands. The primary evolutionary pressures or constraints may have been chemical or thermodynamic, environmental, and ecological.

### 5.1. Pigment absorbance energetics





DRAFT, Kiang, Astrobiology

Here, we examine why the peaks in pigment absorbance are at their particular wavelengths, other than due to being in a light transmittance window.

**Minimum energy requirements at the reaction centers: thermodynamics.**

Conversion of electronic photon energy to chemical energy occurs at the reaction centers. In brief, the energy requirements for this to happen are that the ground state of the primary donor chlorophyll of an RC be at a higher redox potential than the reductant in order to oxidize it, and the photon must be of sufficient energy to excite the primary donor to a sufficiently low redox potential so that it can reduce various intermediates to reduce the final electron acceptor. So, the RC primary donor must straddle the roles of both oxidant in its ground state and reductant in its excited state (Blankenship and Hartman, 1998). In oxygenic photosynthesis (Equation 2), the excitation, electron abstraction, and reductions are achieved in a two-step zig-zag series of potential changes (the "Z-scheme" mentioned earlier), where the reductant $H_2O$ replaces the lost electron from the ground state of Chl a (P680 at 680 nm) in PS II, and PS II supplies the electron to replace that in PS I (P700 at 700 nm), which generates the reduced product eventually used for fixation of $CO_2$. P680 of PS II is at a higher potential than the water, allowing it to be a strong oxidant of water (Blankenship, 2002; Tommos and Babcock, 2000). The potential difference of P680 from water results from the molecular configuration of the remarkable reaction center of PS II. Note that oxidation of water does not depend on the wavelengths of the photons used, but on the midpoint redox potential of the oxygen evolving complex and the reaction center relative to water. In bacterial systems, the potential span afforded by RCs absorbing at 800, 850, and up to 960 nm is smaller and adequate for these other electron donor/acceptor combinations. But, thus, we are still left





DRAFT, Kiang, Astrobiology

with the question, why – on Earth – is P680 at 680 nm, or, more generally, why are the other reaction centers at their exact wavelengths?  If there is no thermodynamic necessity, then perhaps the driving force must be environmental pressure on what light can be harvested.

**Electron transport between light harvesting complexes and reaction center: kinetics.**  In plants and most other photosynthetic organisms, the chlorophylls have peak absorbances at the longest wavelength and, hence, lowest energy compared to the other pigments, which allows the energy cascade via resonance transfer to work.  In PS II, the reaction center's (longest) peak absorption wavelength is at 680 nm (P680), and in PS I it is at 700 nm (P700).  In anoxygenic bacteria, the known reaction centers  are P800 (heliobacteria), P840 (green sulfur bacteria), P870 (purple bacteria, various sulfur and non-sulfur species), P870 (green filamentous, *Chloroflexus aurantiacus*), and P960 (*Blastochloris viridis*, the actual peak is somewhat variable dependent on the solvent or core antenna environment) (Ke, 2001; Blankenship and Prince, 1985).  The core antennae are generally integral to the reaction center complex, but may have slightly different spectral peaks.  The chlorophylls and bacteriochlorophylls also harvest light at a major peak in the blue, but the RCs operate at the red peak.  There are generally about 300 antenna chlorophylls per RC chlorophyll.

In green sulfur, green filamentous, heliobacteria, and in cyanobacteria, the core antennae absorbance spectra peak at shorter wavelengths than the RCs, though there is considerable spectral overlap.  In purple bacteria, the core antennae peak at longer wavelengths than the reaction centers (B1015 in B. viridis whose RC is P960, B890 in several purple bacteria that have P870, and B875 in Rhodobacter sphaeroides, whose RC





DRAFT, Kiang, Astrobiology

is P870; Ke, 2001; Scheer, et al., 2003;  Richard Cogdell and Andrew Gall, personal communication, data in Figure 1b; and Blankenship, 2002).  So, interestingly, light absorbed at wavelengths longer than the reaction centers can be transferred *up* the energy hill to the RCs (Permentier, 2001; Trissl, 1993;  Bernhard and Trissl, 2000).  The transfer mechanism continues to be the subject of debate and research.  In most purple bacteria, the spectral overlap of the RCs and core antennae is sufficient such that thermal variability is enough to prevent exciton energy from being trapped in the core antennae; however, the wavelength separation between the RC and core antenna peaks of *B. viridis* seems to require some other means of energy transfer.  Trissl (1993) proposed a model of the transfer kinetics and trapping times;  assuming a fast thermal equilibration of the excitation energy before the charge separation and constraints on quantum yield, the model seems to explain that the longer wavelength absorption does not affect the trapping time or the quantum yield, and it is profitable for the organism since this allows utilization of the available light. This seems to provide a sensible explanation for the uphill exciton transfer from BChl b at 1013 nm to the RC at 980 nm. Mauzerall and co-workers (Hou, et al., 2001a;  Hou, et al., 2001b;  Boichenko, et al., 2001) quantified, in detail, the entropy changes that occur in PS I, PS II, and cyanobacteria,   and found them to be very small.  Work on kinetics of electron transfer (Trissl, 1993; Trissl, et al., 1999; Bernhard and Trissl, 2000) implies the optimality of arrangements between the reaction centers and light harvesting antennas to ensure efficient transport and trapping of the excitons.

The above theoretical work does not allow for prediction of the wavelength of the reaction centers, but it does offer an explanation of the excitation energy transfer kinetics





DRAFT, Kiang, Astrobiology

between the reaction centers and light harvesting complexes.  The bulk of photosynthetic

activity in *B. viridis* and other purple bacteria is probably driven by mostly shorter-

wavelength photons (David Mauzerall, personal communication), but still, nature finds

ways to harvest the available light.  The light absorbed to 730 nm by oxygenic

photosynthesizers also gets transferred uphill to the P700 reaction center, but this is just

the tail end of the absorption spectrum of chlorophyll in P700 and presumed not to

contribute a large amount to the total photosynthetic activity.  However, Krausz, et al.

(2005) found that there is charge separation in PS II in spinach even over the range 700-

730 nm.  So, uphill energy transfer is not the most productive way to obtain energy, but

nonetheless the phenomenon indicates more means of light harvesting, while resonance

transfer of energy toward the red is the dominant means of light harvesting and energy

trapping.

In general, the literature on light harvesting has focused on excitation energy

transfer kinetics, while the literature on reaction centers has focused on figuring out the

molecular structure and mechanisms, genetic lineage, and molecular evolution (Ferreira,

et al., 2004; McEvoy, et al., 2005;  Xiong, et al., 2000;  Dismukes, et al., 2001).

Meanwhile, theoretical work on the chemical evolution of chlorophyll and the PS II

oxygen-evolving complex is based on ancient ocean chemistry (Mauzerall, 1976;

Dismukes, et al., 2001;  Blankenship, et al., 1998;  Dasgupta, et al., 2004) and provides

energetic constraints on the steps toward the development of PS II.  Little thus far has

been done on solar radiation pressures on the evolution of pigment absorbance spectra, so

we attempt to address such evolutionary drivers here.







DRAFT, Kiang, Astrobiology

**Atmospheric spectral transmittance.**  The available light spectrum, of course, must be the first constraint on reaction center peak absorbance wavelengths.  Numerous studies on diverse photosynthetic organisms show that the radiation absorption spectra of light harvesting pigments matches the spectrum of incident light in the organism's environmental niche.  Examples include:  microbial mats at different water strata (Lengeler, et al., 1999, Eraso and Kaplan, 2001, and Reinhard Bachofen, personal communication); purple and green photosynthetic bacteria in NIR and low light (Blankenship, et al, 1995);  low-light plants (Marschall and Proctor, 2004);  red algae and cyanobacteria absorbing in green wavelengths (Samsonoff and MacColl, 2001);  and, recently discovered, cyanobacteria that perform oxygenic photosynthesis utilizing near-infrared radiation (Chen, et al., 2005).

In the wavelength locations of the onset and plateau of the red edge,  several biological and atmospheric phenomena  occur at both locations.  Details of the charts in Figure 3 are shown in Figure 4 to illustrate more clearly the features around the red edge region.  The bottom of the red edge varies little from 680 nm, which is the peak absorbance wavelength of most of the antenna chorophylls, as well as the primary donor PSII in plants, algae, and cyanobacteria. .  However, depending on measurement precision or perhaps cell structures, the onset can be as low as 670 nm (lichen, Licedea-1 in Fig 4c) and does not appear to go beyond 700 nm, where some of the long-wavelength forms of antenna chlorophylls absorb.  Also, the Chl a primary donor of PS I has its peak absorbance at 700 nm.  Significant phenomena occur in the red edge region (Krausz, et al., 2005, called it "spectral congestion" with regard to an even more detailed structural





# DRAFT, Kiang, Astrobiology

breakdown than summarized here) at the following wavelengths: 650-670 nm, where core

antenna minor sub-bands occur for PS I and PS II; 678.5 nm, the main sub-band for PS

II; 680 nm, the location of the primary donor P680 for PS II; 682 nm, the main sub-band

for PS I; and 700 nm, the primary donor P700 for PS I. The main sub-bands are actually

where most of the light harvesting of the core antenna occurs. From these data, it appears

that the maximum absorbance at the foot of the red edge rarely departs from the 678.5 nm

sub-band in organisms that utilize PS II.

Perhaps the most significant spectral feature to observe is that the maximum

photon flux density at the Earth's surface occurs at 685 nm, just before a drop in

atmospheric transmittance due to oxygen at 687.5 nm. The $O_3$ Chappuis band (500-700

nm) shifts the Sun's photon spectral flux density from its top-of-the-atmosphere peak at

600 nm to 685 nm at the Earth's surface, which may partially explain why chlorophyll

favors the red rather than the green. (Note that, at the Earth's surface, the incident energy

flux peak is spread over 450-490 nm, in the green). Thus, it appears that the peak

absorbance at the foot of the red edge is an adaptation to harvesting light in the

atmospheric transmittance window with the most abundant photon flux, and the peak is at

the most red-shifted limit of that window to afford exciton transfer from accessory

pigments at the shorter wavelengths in that window. The long wavelength limit of this

window is due to both the solar spectrum and the presence of oxygen, the very product of

photosynthesis. Note that, if the surface incident light spectrum is viewed in terms of

energy flux rather than photon flux, the peak flux is around 480-490 nm, which is in the

blue-green; since photosynthesis counts photons, not total energy, it is the peak photon

flux that is favored by pigments. On the other hand, of course, if photosynthesis evolved





DRAFT, Kiang, Astrobiology

originally under water prior to atmospheric oxygen build-up, then there is no good reason

for the oxygen band to have supplied any evolutionary pressure on the reaction center

peaks, unless such selective pressure continued to occur on near-surface organisms.

However, this may explain why PS I and PS II of green cyanobacteria and green algae

had the advantage in giving rise to land plants.

The NIR end of the red edge is clearly confined to wavelengths between the

oxygen A-band at 761 nm and the bluest side of the water band at about 718 nm.  Only

the snow algae have an NIR end blue-shifted from this 718 nm.  One might expect

organisms under ice or water  to have spectral characteristics adapted to these media in

contrast to air.  However, there is no distinctive feature for ice absorption or reflectance

blue-ward of the water band, and for organisms under water, there is not clearly a tight

relation with the water absorption band in our data.  Figure 6b shows our one anoxygenic

example together with the irradiance through water.  It may be that the peak absorption

wavelength of this purple bacteria is related to the water band at ~810-840 nm.  More

whole-organism reflectance data are needed to draw any firm conclusions, but

tentatively, we hypothesize that the NIR end has evolved in response to major absorbance

bands of the air or water medium of the photosynthesizer.

Purple bacteria have a starkly different reflectance spectrum, with our one

example showing an "NIR edge" (perhaps better called an "NIR slope") rather than a red

edge, starting at 837 nm.  The bacteriochlorophylls for the purple bacteria and green non-

sulfur bacteria result in primary donor or reaction center wavelengths at 840 to 960 nm.

Unfortunately, few single-colony or whole-organism reflectance spectra have been

measured of anoxygenic bacteria.The data here were measured only over 400-900 nm,so





DRAFT, Kiang, Astrobiology

we cannot examine other NIR features or the spectrum of species like *B. viridis* that harvest light at longer wavelengths.

### 5.2. NIR scattering

As seen in Figures 4-7, not all photosynthetic organisms exhibit the same degree of NIR scattering. Why does the NIR reflectance vary, whether due to environmental adaptations, physiological status, or other unexplained evidence?

**Morphological adaptations to light and climate.** It is well known that morphological adaptation to different climate limits and light levels will influence spectral characteristics due to differences in leaf thickness or canopy density. These quantities affect the boundary layer conductance at the leaf surface and light penetration into the leaf (where the leaf boundary layer is the gas diffusive layer of air at the surface). Thus, cold, dry environments favor needleleaf plants, wet, temperate environments favor broadleaf species, and hot, dry environments favor succulents with low surface-to-volume ratios (Holdridge, 1967; Larcher, 1995; Schuepp, 1993; Foley, et al., 1996). Highly sunlit leaves will be thicker to allow for a greater absorption cross section of photosynthetically active radiation (Reich, et al., 1997; Kull, 2002); therefore, the visible/NIR contrast will be stronger for high light-adapted leaves. In addition, leaf surface characteristics, such as hairs or trichomes, specularity of the surface, and surface waxes, can affect the overall reflectance; hairs help increase albedo in hotter, brighter environments, while waxes may serve to absorb high UV in arctic environments.





DRAFT, Kiang, Astrobiology

**Compactness of cells.**  A lower NIR reflectance may result from a lack of air spaces around more compact cells.  For example, loss of cell turgor during water stress could reduce air-cell wall interfaces for NIR scattering.  Lichens have a dense structure in a fungal cortex that overlies their cyanobacterial layer, and therefore, they are not highly reflective in the NIR.  Conifer species have a dense mesophyll structure, and hence, their leaves tend to be darker in the NIR than those of broadleaf plants.  In contrast, *Sphagnum* moss increase rather than decrease NIR reflectance when dried (dried spectra not shown), because their means of water supply to the plant's capitula is not through internal conducting cells but  through precipitation or capillary rise only (Harris, et al., 2005). Abundant hyaline cells provide a large water holding capacity in Sphagnum, but their drying results in structural changes much different from that of vascular plant leaf mesophyll cells.  *Sphagnum magellicanum* (moss Figure 4c) has a very low-sloping red edge compared to the other mosses, due to more tightly bunched capitula.

**Energy balance.**  Although the NIR reflectance must play some role in an organism's energy balance, it is not clear how important this is in the organism's survival and evolution of its spectral signature.  In snow algae, Gorton, et al. (2001) found a substantial NIR absorbance of the outer membrane, which could possibly afford a more favorable energy balance for the algae.  On the other hand, as noted above, the NIR reflectance decreases in desiccated plant leaves but increases in dried mosses.  Lichens, in both cold arctic and hot tropical environments, exhibit very low scattering (reflectance as well as transmittance) in the NIR (transmittance <15%, Bechtel, et al., 2002).  Aquatic plants and algae and cyanobacteria that grow under water have no clear need for an NIR reflectance to control their energy balance.  The NIR reflectance can be found to vary by





DRAFT, Kiang, Astrobiology

the same degree in diverse ecosystems across different environmental gradients, such that it is not straightforward to draw a conclusion about the NIR reflectance's role in an organism's survival or competitive advantage.  Undoubtedly, it is important to some organisms, but there are not yet enough data to determine consistent trends.  More studies are needed of the relation between the energy balance of photosynthetic organisms, their climatic limits, and their spectral reflectance signatures.

   **Cell wall refractive index and light transmission.** The cell wall composition of organisms, of course, must have a definite functional role in exchange of gas and fluid, structural support (or not), and transmission of light, and it determines the spectral refractive index, as mentioned earlier.  In Table 1, it can be seen that there are some differences between taxa.  Terrestrial plant cell walls contain cellulose, lignin, polysaccharides, and protein, whereas aquatic plants contain little or no lignin, since it is not necessary for support.  Algal cell walls are predominantly cellulose; dinoflagellates, which are the photosynthetic symbionts in corals and are responsible for algal blooms known as red tides, also have a theca, an armor-like set of plates beneath the plasma membrane (Larkum and Vesk, 2003;  Evitt, 1985).  Cyanobacteria cell walls contain murein.  Unfortunately, the table is incomplete, since there are few data on cell wall compositions for other photosynthetic organisms.  The cell wall composition of purple bacteria might not have very different refractive properties from that of other photosynthesizers, so long as the NIR radiation can be transmitted into the cell.

**6.  Conclusions**





DRAFT, Kiang, Astrobiology

To summarize, the full reflectance spectrum of a photosynthetic organism is the expression of both molecular and macrostructural properties: 1) the absorbance spectra of the light-harvesting complex, the core antenna complex, and the reaction centers, 2) cell membrane and cell wall refractive properties, and 3) whole-organism structural impacts on light scattering. How the above properties evolved is not well known, but a number of environmental pressures and molecular constraints play a role: 1) the thermodynamics of light harvesting and exciton transfer kinetics, 2) the redox potential requirements for oxidation of the electron donor and reduction of $CO_2$, 3) adaptation to available resources (light spectrum, nutrients, electron donor), which is not covered in detail here, 4) adaptation for protection against environmental harm (UV radiation, temperature, chemical toxicity, e.g. levels of pH, $O_2$, other).

Electron abstraction from the reductant, such as $H_2O$, $H_2S$, or $FeOH^+$, does not depend on the wavelength of the photon but on the redox potential of the biochemical molecule. The excitation of the reaction center chlorophyll to a sufficiently energetic state does depend on the photon energy, and multi-photosystem pathways could theoretically utilize more photons at longer wavelengths to evolve $O_2$ and fix carbon. A long wavelength limit might be 1100 nm, a threshold between optical and thermal or vivartional. The ability to perform electronic transitions, however, depends on the molecule, and more research is needed to define strict thermodynamic contraints. The kinetics of exciton transfer require sufficent proximity between light harvesting pigments and the reaction center for efficient trapping of the excitons, such that it is more energetically favorable for the RC to absorb at longer wavelengths; however, uphill





# DRAFT, Kiang, Astrobiology

electron transfer does occur.  The spectrum of available light ultimately limits pigment

absorption spectra and productivity.

For the full reflectance spectrum of the organism, we observed, on Earth, that the

"red edge" is nearly ubiquitous among oxygenic photosynthesizers, but it is weak or

negligible in lichens. For cyanobacteria, more data are needed.  Although it is fairly well

understood how the NIR reflectance varies due to morphology, the selective role of the

NIR reflectance is not well understood.  Thus far there is anecdotal evidence with regard

to the organism's energy balance (and one study on crop leaves by Aboukhaled, 1966)

and not enough data on the diversity of cell membrane and cell wall compositions.  For

anoxygenic photosynthetic bacteria that have their reaction centers in the NIR, the one

example for purple bacteria shows no red edge, as it would not make sense for the

organism to scatter light in the wavelengths that it utilizes.  The green sulfur bacteria,

oddly, appear to exhibit a bit of red edge, so more measurements are needed to confirm

whether this is the general case.  The spectra we assembled show striking trends in the

NIR end among organisms, with lichens, algae, and mosses most blue-shifted, terrestrial

plants most red-shifted, and aquatic plants in the middle.  More data on bacteria and algae

across environmental gradients, along with consistent measurement of ambient conditions

and assessment of the organisms physiological status, are needed to confirm the strength

of these trends.

The Sun's radiation spectrum and the spectral transmittance of the atmosphere

and water environments are the most important selective pressures on critical points in

the pigment spectra of photosynthetic organisms.  Atmospheric oxygen may have altered

the atmospheric transmittance spectrum enough to favor PS I and PS II absorbance in the





DRAFT, Kiang, Astrobiology

red.  Meanwhile, the example of lichens indicates that not all photosynthesizers will necessarily have a steeply contrasting reflectance where pigments do not absorb and high NIR reflectance is not clearly an energy balance adaptation.  So, we cannot conclude yet how a whole organism's reflectance spectrum, besides the pigment spectra, is a function of environmental adaptation.  We do not have enough data on bacteria to draw conclusions about their reflectance properties, but the one purple bacteria example indicates that shifted spectral signatures are possible for organisms using anoxygenic photosystems.  Resonance transfer and exciton transfer kinetics appear to work in concert with the available light spectrum to constrain the peak absorbance wavelength of the core antenna and reaction centers. Given the ability of organisms to transfer light energy both downhill and uphill to the reaction centers, it may be sufficient just to search for a pigment signature within particular atmospheric transmittance windows.

We can, therefore, propose the following candidates for photosynthetic pigment peak absorbance wavelengths:

a.  the wavelength of peak incident photon flux within a radiation transmittance window, as the main environmental pressure;

b.  the longest wavelength within a radiation window for core antenna or reaction center pigments, due to the resonance transfer of excitation energy and an energy funneling effect from shorter to longer wavelengths;





DRAFT, Kiang, Astrobiology

    c. the shortest wavelengths within an atmospheric window for accessory

pigments, also due to resonance transfer.

    These hypotheses assume some optimality principle, where the peak photon flux

wavelength is energetically and ecologically most favorable for survival, and organisms

will have adapted to this.  It may be that accidents, inertia, or the slow stages of evolution

will not yield the optimum spectral signature for photosynthesizers at the time we

observe them.  For example, on Earth, the inefficiency of Rubisco (the carbon fixing

enzyme on which all photosynthesis depends and which sometimes is rendered useless

for carbon fixation, because it also functions as an oxygenase) is the subject of much

lament and agricultural biotechnological research (Parry, et al., 2003).  Plant reflectance

of green light is often considered a similarly sub-optimal feature of plants, as there

appears a spectral mismatch between solar radiation at the Earth's surface and the

absorption peaks (440 and 680 nm) of chlorophyll (Raven and Wolstencroft, 2002).

Even with our rules above for pigment properties, there appears to be wasted green light.

However, given that algae and cyanobacteria utilize phycobilins to harvest green light,

the low harvesting of green light in land plants has been thought by some to be an

inefficiency due to an evolutionary "lock-in" from the lineage of green algae.  However,

above-ground plants pigments do absorb green light, just in a lesser ratio to other colors,

and they often experience too much light (or are limited by other resources), hence the

need for quenching by carotenoids.  The non-light harvesting pigment anthocyanin,

which accumulates at the surface of shade-adapted leaves and makes them red, may

provide photo-protection under high light by shading Chl b in chloroplasts from green





DRAFT, Kiang, Astrobiology

light (Gould, et al., 2002;  Pietrini, et al., 2002).  So, there may be no selective advantage

to absorbing more green light.  That all land plants are descended from the green algae

may very much be because their pigment combinations provided the selective advantage

for life on land after the build-up of atmospheric $O_2$, which shifted the surface spectral

photon flux from a peak at 600 nm to 685 nm.  So, given some caveats about

evolutionary optimality with regard to light resource constraints, we propose the above

rules for wavelengths of peak photosynthetic pigment radiation absorbance, dependent on

the spectral photon flux density at the surface of a planet.

To explain whole-organism spectral reflectance and the steepness of absorbance

peaks, more studies are needed of: bacteria spectral properties;  the role of pigment-

protein complexes in altering pigment absorbance spectra; thermodynamic limits of light

harvesting and redox biochemistry; field studies and modeling of organism radiative

transfer, growth, and energy balances within their respective light environments; and

further speculation on the environment of the early Earth.

# Acknowledgments

Numerous people kindly contributed data and valuable advice for this paper.  We

are greatly obliged to Niels-Ulrik Frigaard, Richard Cogdell, and Andrew Gall for

pigment data and valuable comments;  Michael Eastwood, Robert Green, and Scott Nolte

of the JPL/AVIRIS lab for the use of their spectroradiometer;  Mike Schaadt and the

Cabrillo Marine Aquarium for providing marine algae samples.  Holly Gorton for snow

algae data;  Suzanne Fyfe for seagrass spectra;  Robert Bryant and Angela Harris for

moss spectra;  Greg Asner for lichen spectra from Hawaii;  Reinhard Bachofen for





# DRAFT, Kiang, Astrobiology

microbial mat bacteria data; Brian Cairns, Judith Lean, and Andrew Lacis for solar spectral photon flux densities; Dennis Clark for Hawaii buoy radiation data; Stephane Jacquemoud for the use of the PROSPECT model; Warwick Hillier, Yongqin Jiao, and Elmars Krausz for very helpful explanations about photochemistry. Thanks are also due to John Scalo, Norm Sleep, and Jim Kasting for many lively discussions and helpful references. We also are grateful to the many people who have made their datasets available on-line and who are cited in this paper. Finally, we greatly appreciate the very helpful comments of two anonymous reviewers. M.C. thanks NASA for supporting his participation in this work through JPL contract 1234394 with UC Berkeley. Govindjee thanks the Dept. of Plant Biology of the University of Illinois for office support. N.Y.K also thanks NASA and James Hansen for supporting this work.






# DRAFT, Kiang, Astrobiology

DRAFT, Kiang, Astrobiology

DRAFT, Kiang, Astrobiology

DRAFT, Kiang, Astrobiology

DRAFT, Kiang, Astrobiology

# DRAFT, Kiang, Astrobiology

DRAFT, Kiang, Astrobiology

*For Peer Review*





# Table captions

Table 1.  Range of phototrophic organisms on Earth.   Bacteria compiled by Janet Siefert.  Additional carbon sources and electron donors from Overmann and Garcia-Pichel  (2000), Decker, et.al. (2005), Blankenship, et al. (1995), Eraso and Kaplan (2001), Miller, et al. (2005).  Algae from Larkum, et.al, (2003).  Plants from (Nobel, 1999).  Abundance and productivity numbers from Reeburgh (1997) and Cramer, et al. (1999).  Bacterial abundance and productivity are autotrophic only, primarily marine cyanobacteria, estimated from Whitman, et al. (1998).





Table 1. Range of phototrophic organisms on Earth. Bacteria compiled by Janet Siefert. Additional carbon sources and electron donors from Overman and Garcia-Pichel (2000), Decker, et.al. (2005), Blankenship, et al. (1995), Eraso and Kaplan (2001), Miller, et al. (2005). Algae from Larkum, et al, (2003). Plants from (Nobel, 1999). Abundance and productivity numbers from Reeburgh (1997) and Cramer, et al. (1999). Bacterial abundance and productivity are autotrophic only, estimated from Whitman, et al. (1998).

| Appeared | Taxon | Growth mode/form | Cell wall | RC | Pigments | e-donor | C source* | Products | Niche | Global Abundance (Pg-C) | Gross Productivity (Pg-C/yr) |
|---|---|---|---|---|---|---|---|---|---|---|---|
| | BACTERIA | unicellular | | | | | | | | 0.06 | 51 |
| 3.8 Ga | Anoxygenic | | | | | | | | | | |
| | Green non-sulfur filamentous | anoxygenic photoorganoheterotroph aerobic chemoorganoheterotroph | | Type II | BChl a/c/ and/ or d/e +car | sulfide | orgC; CO2 | S | dense microbial mats in hot springs often in association with cyano, thermophilic; floculent surface layer in alkaline springs; Chloroflexus: max ~70 Celsius, cannot fix N, resistant to UV | | |
| | Green sulfur bacteria | anoxygenic photolithoautotroph | | Type I | BChl a + c/d/e +car | sulfide, reduced S, $H_2$, Fe | orgC: acetate, propionate, pyruvate; $CO_2$ | sulfate | non-thermal aquatic ecosystems, hot springs, max 55-56 Celsius (Chlorobium tepidum) | | |
| 3.8 Ga | Purple bacteria | aerobic and anoxgenic heterotrophs and anoxygenic autotroph | | Type II | BChl a/b +car | inorg&orgC S, sulfate, sulfide, sulfite, $H_2$, Fe | orgC, $CO_2$ | S, sulfate, $CO2$ | fresh and marine waters, eutrophic marine, hot springs, anoxic aquatic sediments; max >50 Celsius (Chromatium tepidum) | | |
| | Heliobacteria | anoxygenic photoorganoheterotroph | | Type I | BChl g+car | sulfide, reduced S, sulfate | pyruvate, ethanol, lactate, acetate, and butyrate | ? | soil, dry paddy fields, occasionally lakeshore muds, hot springs; resistant to UV, fix N; survive at least to 42 Celsius | | |
| | Halobacteria (not actual photosynthesis) | aerobic chemoorganoheterotroph | C5 isoprenoid chains attached to glycerol by ether linkages | bacterio-rhodopsin | bacterio-rhodopsin | N/A | orgC | ? | salt crusts in marine salterns, saline lakes, evaporites, ~4M NaCl | | |
| 3.6-2.3 Ga | Oxygenic | | | | | | | | | | |
| 3.6-2.3 Ga | Cyanobacteria | oxygenic photolithoautotroph | murein | Type I & II | Chl a/b/c/d +PBS+car | $H_2O$, S | $CO_2$ | $O_2$ | everywhere, -15 to +75°C | ? | ? |
| 1.2 Ga | ALGAE | eukaryotes, autotrophs, unicellular, multicellular | | Type I & II | | $H_2O$, other? | $CO_2$ | $O_2$ | fresh and marine waters, snow | 2 | ~100 |
| 1.2 Ga | Rhodophytes (red algae) | " | cellulose, xylan | | Chl a + PBS + car | | | | min observed PAR flux 0.01 micromol/m2/s (Littler, et.al., 1986) | | |
| | Chromophytes (incl. brown algae) | " | cellulose; diatoms: silica, alginate | | Chl a/c + car | | | | | | |
| 750 Ma | Chlorophytes (green algae) | | cellulose | | Chl a/b/c +PBS +car | | | | | | |
| | Lichens crustose, squamose, foliose, fruticose | symbiosis of fungus and cyanobacteria/algae | | | Chl a/b+car | $H_2O$ | | | rock outcrops, vegetation surfaces | ? | ? |
| | PLANTS | leaves, stems, roots | cellulose, lignin, polysaccharides (e.g. pectin), protein, water, calcium | Type I & II | Chl a/b+car | $H_2O$ | $CO_2$ | $O_2$, VOCs | min observed PAR ~ 3 micromol/m2/s (0.7 W/m2) | 550-680 | ~90-120 |
| 460 Ma | Bryophytes | non-vascular | | | | | | | moist land environments | | |
| | Mosses Liverworts | Sphagnum, Acrocarpus, Pleurocarpus | | | | | | | | | |
| | Vascular plants | broadleaf, needleaf, herbaceous, succulent | | | | | | | aquatic to desert environments | | |
| | Aquatic plants | | no lignin | | | | $CO_2$, $HCO_3^-$ | | | | |
| | C3 pathway | | | | | | | | | | |
| 144 Ma | Flowering plants | | | | | | | | | | |
| 70-55 Ma | CAM | | | | | | | | | | |
| 20-35 Ma | C4 pathway | | | | | | | | | | |

Ga - billion years ago, Ma - million years ago, RC-reaction center, PBS-phycobilisomes, car-carotenoids, VOC-volatile organic compound. *C source related only to the photosynthetic process (disregards carnivorous plants).





# Figure Captions

Figure 1.  Electron transport pathways of photosynthesis, with midpoint redox potentials of ground and excited states of the reaction centers, of biochemical intermediates and reduced products.  Shown are photosystems for purple bacteria, green sulfur bacteria, and oxygenic photosynthesis.

Figure 2.

a) Solar spectral photon flux densities at the top of the Earth's atmosphere and at the Earth's surface, and estimated *in vivo* absorption spectra of photosynthetic pigments of plants and algae.  *Sources*:  Modeled photon flux densities from the following: Top-of-the-atmosphere (TOA) irradiance: 150-200 nm, Andrew Lacis, NASA Goddard Institute for Space Studies (GISS);  200-400 nm, Judith Lean (Naval Research Laboratory);  400-2500 nm, Brian Cairns, NASA GISS.  Surface irradiance:  200-400 nm, J. Lean (Lean and Rind, 1998);  400-2500 nm, Brian Cairns.  Hawaii buoy measurements from Dennis Clark (NOAA).  Chlorophyll a and Chlorophyll b absorbance measurements, made by Junzhong Li (H. Du and coworkers, 1998), *in vitro*, were shifted in wavelengths to match *in vivo* peaks, and absorbances were normalized to between 0 and 1.  Carotenoid absorption spectra are estimated *in vivo* absorption spectra in green algae (Govindjee, 1960).  Phycoerythrin and phycocyanin absorption spectra are unpublished absorption spectra from Govindjee's laboratory, and from Ke (2001).  Chlorophyll a





fluorescence spectrum, from spinach chloroplasts, is from Govindjee and Yang

(1966).  Pigments, measurement method, and sources are listed in Appendix A1.

b) Solar spectral photon flux densities at the top of the Earth's atmosphere, at the

Earth's surface, at 5 cm depth in pure water, and at 10 cm depth of water with an

arbitrary concentration of brown algae;  algae and bacteria pigment absorbance

spectra.  *Sources:*  Top-of-the-atmosphere (TOA) and surface incident radation

same as Figure 2a.  Water spectral absorption coefficient:  200-380 nm, Segelstein

(1981); 380-640 nm, Sogandares, et al. (1997);  640-2500 nm, Kou, et al., (1993).

Algae (brown, kelp, *Macrocystis pyrifera*) absorption coefficient from reflectance

spectrum measured (in lab, in air) by N.Y. Kiang with ASD FieldSpec

spectroradiometer (instrument from JPL/AVIRIS Lab). Bacteriochlorophyll

pigment absorbance spectra are all *in vivo* in intact membranes, including

carotenoids.  BChl a (*Rhodobacter sphaeroides*) and BChl b (*Blastochloris

viridis*) spectra from Richard Cogdell and Andrew Gall.  BChl c, d, and e spectra

from green sulfur bacteria (Frigaard, et al., 2004). Pigments, measurement

method, and sources are listed in Appendix A1.

c) Solar spectral photon flux densities at the top of the Earth's atmosphere (TOA)

and at the Earth's surface, with reflectance spectra of terrestrial plants, moss, and

lichen (source:  Clark, et al., 2003), and $O_2$ and $H_2O$ absorbance lines.





Figure 3. Modeled reflectance spectra of a generalized plant leaf, from the model PROSPECT (Jacquemoud and Baret, 1990). Variations in reflectance due to: a) structure, b) water content, c) Chl a and b content, and d) carbon content.

Figure 4. Reflectance spectra of different photosynthetic organisms, minerals, and non-photosynthetic organic matter in 0.2-2.4 micron range. The vertical dotted line in all the plots is at 0.761 micron, corresponding to an oxygen absorption line. *Sources*: a) Land plants: Clark, et al. (2003). Surfgrass, aquatic plant: N.Y. Kiang. . b) Aquatic plants: Fyfe, et al. (2003); surfgrass: N.Y. Kiang. c) Mosses: Lang, et al. (2002), Harris, et al., (2005). d) Lichens: *Cladina* and *Sterocaulon*, courtesy of Greg Asner; *Acarospora, Licedea, Xanthoparmelia, and Xanthoria,* Clark, et al. (2003). e) Algae: N. Kiang, snow algae: Gorton, et al. (2001). f) Bacteria in a microbial mat: from Reinhard Bachofen in Wiggli, et al. (1999). g) Minerals and golden dry grass: Clark, et al. (2003). Human skin: N.Y. Kiang. Species names and instruments used are listed in Appendix A2.

Figure 5. Details of Figure 4. Reflectance spectra of different photosynthetic organisms, minerals, and non-photosynthetic organic matter over 0.4-0.9 microns. *Sources*: Same as for Figure 4. a) Land plants: Clark, et al. (2003). Surfgrass,





aquatic plant: N.Y. Kiang. b) Aquatic plants: Fyfe, et al. (2003); surfgrass: N.

Y. Kiang. c) Mosses: Lang, et al. (2002) ), Harris, et al., (2005). d) Lichens:

Clark, et al. (2003). e) Algae: N.Y. Kiang, snow algae: Gorton, et al. (2001). f)

Bacteria in a microbial mat: from Reinhard Bachofen in Wiggli, et al. (1999).

Figure 6. Calculation of the NIR end wavelength, aquatic plant, *Posidiana*

*australis*. Vertical solid thin line is location of 3$^{rd}$ derivative maximum and NIR

end of the spectral reflectance (data courtesy of Susan Fyfe).

Figure 7. Scatterplot of wavelengths of the red edge inflection point ("edge

wavelengths") and the NIR end ("plateau wavelengths") for organisms in Figures

4 and 5. Vertical axis has no scale, but points are dithered vertically simply to

show their horizontal spread.





# Electron Transport Pathways of PS

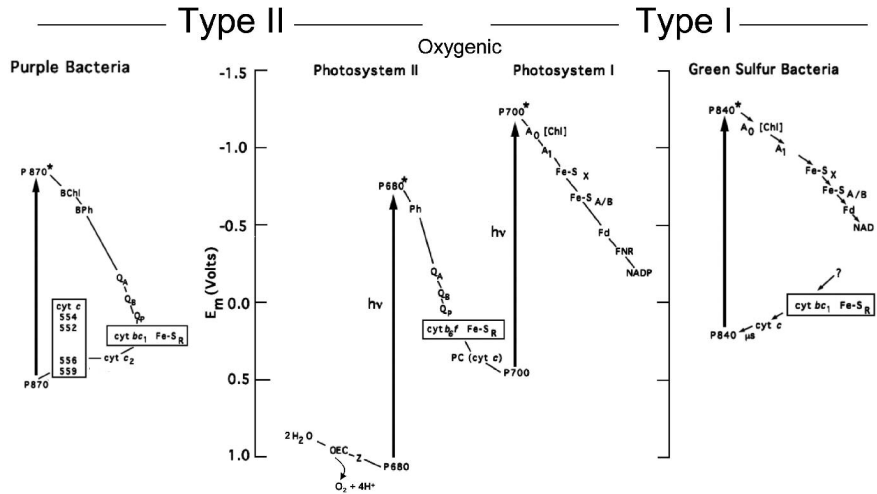

**Figure 1– Electron transport pathways of photosynthesis, with midpoint redox potentials of ground and excited states of the reaction centers, of biochemical intermediates and reduced products. Shown are photosystems for purple bacteria, green sulfur bacteria, and oxygenic photosynthesis.**





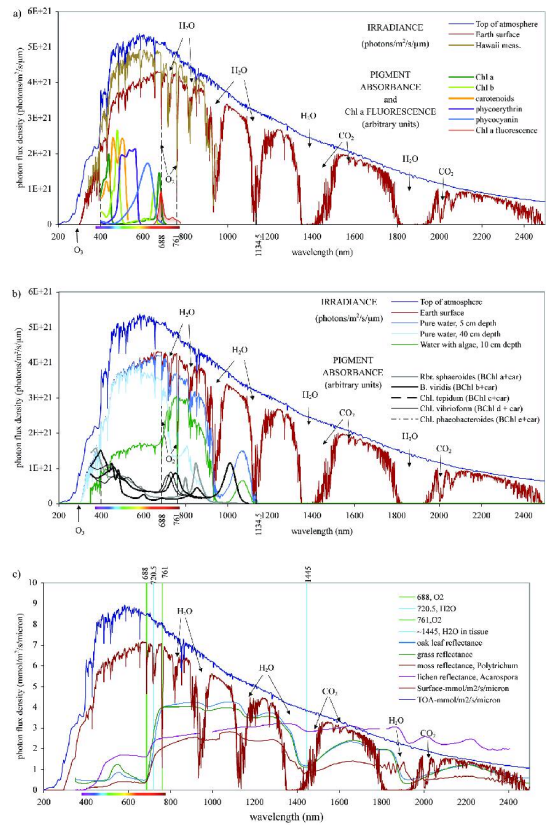

Figure 2

**Figure 2- a) Solar spectral photon flux densities at the top of the Earth's atmosphere and at the Earth's surface, and estimated in vivo absorption spectra of photosynthetic pigments of plants and algae. Sources: Modeled photon flux densities from the following: Top-of-the-atmosphere (TOA) irradiance: 150-200 nm, Andrew Lacis, NASA Goddard Institute for Space Studies (GISS); 200-400 nm, Judith Lean (Naval Research Laboratory); 400-2500 nm, Brian Cairns, NASA GISS. Surface irradiance: 200-400 nm, J. Lean (Lean and Rind, 1998); 400-2500 nm, Brian Cairns. Hawaii buoy measurements from Dennis Clark (NOAA). Chlorophyll a and Chlorophyll b absorbance measurements, made by Junzhong Li (H. Du and coworkers, 1998), in vitro, were shifted in wavelengths to match in vivo peaks, and absorbances were normalized to between 0 and 1. Carotenoid absorption spectra are estimated in vivo absorption spectra in green algae (Govindjee, 1960). Phycoerythrin and phycocyanin absorption spectra are unpublished absorption**





spectra from Govindjee's laboratory, and from Ke (2001). Chlorophyll a fluorescence spectrum, from spinach chloroplasts, is from Govindjee and Yang (1966). Pigments, measurement method, and sources are listed in Appendix A1. b) Solar spectral photon flux densities at the top of the Earth's atmosphere, at the Earth's surface, at 5 cm depth in pure water, and at 10 cm depth of water with an arbitrary concentration of brown algae; algae and bacteria pigment absorbance spectra. Sources: Top-of-the-atmosphere (TOA) and surface incident radation same as Figure 2a. Water spectral absorption coefficient: 200-380 nm, Segelstein (1981); 380-640 nm, Sogandares, et al. (1997); 640-2500 nm, Kou, et al., (1993). Algae (brown, kelp, Macrocystis pyrifera) absorption coefficient from reflectance spectrum measured (in lab, in air) by N.Y. Kiang with ASD FieldSpec spectroradiometer (instrument from JPL/AVIRIS Lab). Bacteriochlorophyll pigment absorbance spectra are all in vivo in intact membranes, including carotenoids. BChl a (Rhodobacter sphaeroides) and BChl b (Blastochloris viridis) spectra from Richard Cogdell and Andrew Gall. BChl c, d, and e spectra from green sulfur bacteria (Frigaard, et al., 2004). Pigments, measurement method, and sources are listed in Appendix A1. c) Solar spectral photon flux densities at the top of the Earth's atmosphere (TOA) and at the Earth's surface, with reflectance spectra of terrestrial plants, moss, and lichen (source: Clark, et al., 2003), and $O_2$ and $H_2O$ absorbance lines.





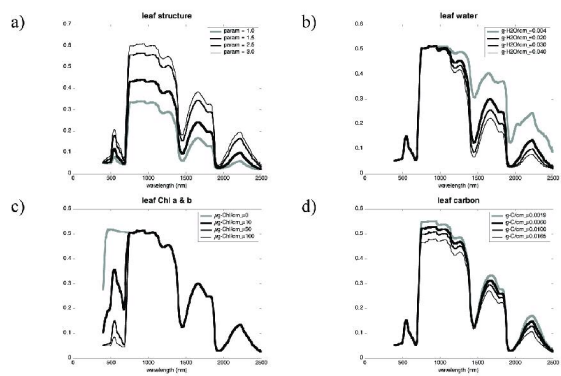

Figure 3.  Modeled reflectance spectra of a generalized plant leaf, from the model
PROSPECT (Jacquemoud and Baret, 1990).  Variations in reflectance due to: a)
structure, b) water content, c) Chl a and b content, and c) carbon content.

**Figure 3- Modeled reflectance spectra of a generalized plant leaf, from the model
PROSPECT (Jacquemoud and Baret, 1990). Variations in reflectance due to: a) structure,
b) water content, c) Chl a and b content, and d) carbon content.**





1
2
3
4
5
6
7
8
9
10
11
12
13
14
15
16
17
18
19
20
21
22
23
24
25
26
27
28
29
30
31
32
33
34
35
36
37
38
39
40
41
42
43
44
45
46
47
48
49
50
51
52
53
54
55
56
57
58
59
60

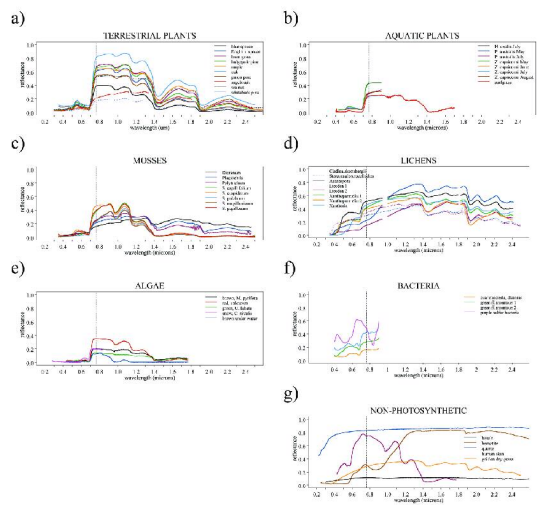

Figure 4

**Figure 4– Reflectance spectra of different photosynthetic organisms, minerals, and non-photosynthetic organic matter in 0.2-2.4 micron range. The vertical dotted line in all the plots is at 0.761 micron, corresponding to an oxygen absorption line. Sources: a) Land plants: Clark, et al. (2003). Surfgrass, aquatic plant: N.Y. Kiang. . b) Aquatic plants: Fyfe, et al. (2003); surfgrass: N.Y. Kiang. c) Mosses: Lang, et al. (2002), Harris, et al., (2005). d) Lichens: *Cladina* and *Sterocaulon*, courtesy of Greg Asner; *Acarospora, Licedea, Xanthoparmelia*, and *Xanthoria*, Clark, et al. (2003). e) Algae: N. Kiang, snow algae: Gorton, et al. (2001). f) Bacteria in a microbial mat: from Reinhard Bachofen in Wiggli, et al. (1999). g) Minerals and golden dry grass: Clark, et al. (2003). Human skin: N.Y. Kiang. Species names and instruments used are listed in Appendix A2.**





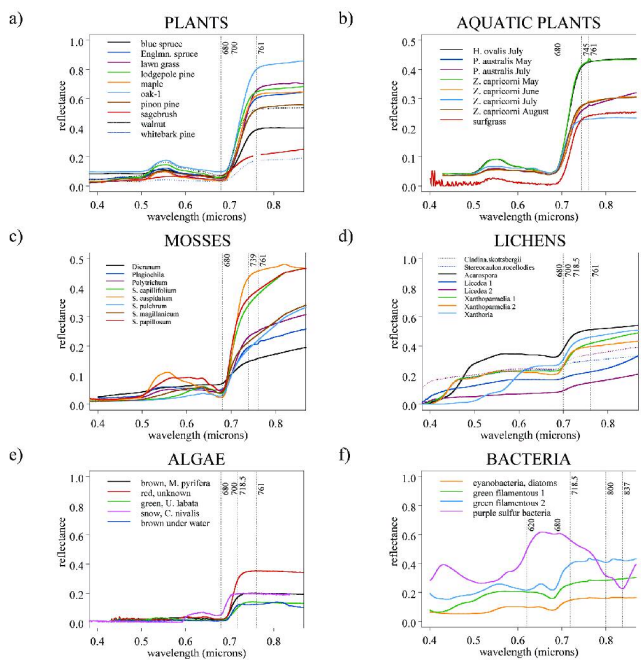

Figure 5.

**Figure 5- Details of Figure 4. Reflectance spectra of different photosynthetic organisms, minerals, and non-photosynthetic organic matter over 0.4-0.9 microns. Sources: Same as for Figure 4. a) Land plants: Clark, et al. (2003). Surfgrass, aquatic plant: N.Y. Kiang. b) Aquatic plants: Fyfe, et al. (2003); surfgrass: N. Y. Kiang. c) Mosses: Lang, et al. (2002) ), Harris, et al., (2005). d) Lichens: Clark, et al. (2003). e) Algae: N.Y. Kiang, snow algae: Gorton, et al. (2001). f) Bacteria in a microbial mat: from Reinhard Bachofen in Wiggli, et al. (1999).**





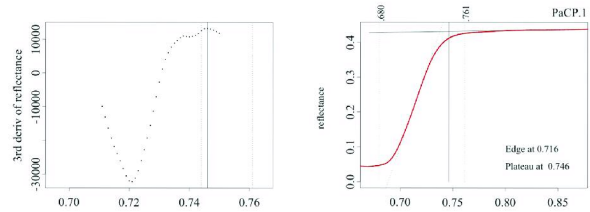

Figure 6. Example of NIR end calculation for an aquatic plant, *Posidonia australis* (data courtesy of Susan Fyfe).

**Figure 6– Calculation of the NIR end wavelength, aquatic plant, *Posidiana australis*. Vertical solid thin line is location of 3rd derivative maximum and NIR end of the spectral reflectance (data courtesy of Susan Fyfe).**





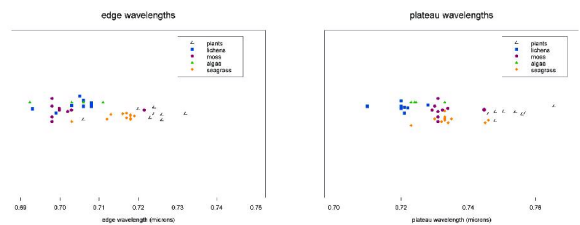

Figure 7. Scatterplot of wavelengths of the red edge inflection point ("edge wavelengths") and the NIR end ("plateau wavelengths") for organisms in Figures 3 and 4. Vertical axis has no scale, but points are dithered vertically simply to show their horizontal spread.

**Figure 7– Scatterplot of wavelengths of the red edge inflection point (    edge wavelengths    ) and the NIR end (    plateau wavelengths    ) for organisms in Figures 4 and 5. Vertical axis has no scale, but points are dithered vertically simply to show their horizontal spread.**





**Appendix A1.** Photosynthetic pigment absorbance spectra

| Pigment | Object measured | Source |
|---|---|---|
| Chl a | *in vitro* spectra were stretched/shifted in wavelengths to match *in vivo* peaks by linear transformation | *In vitro* spectra from Junzhong Li (H. Du and coworkers, 1998) |
| Chl b | *in vitro* spectra were stretched/shifted in wavelengths to match *in vivo* peaks by linear transformation | *In vitro* spectra from Junzhong Li (H. Du and coworkers, 1998) |
| BChl a | intact membranes in *Rhodobacter sphaeroides* | Richard Cogdell and Andrew Gall (pers. comm.) |
| BChl b | intact membranes in *Blastochloris viridis* | Richard Cogdell and Andrew Gall (pers. comm..) |
| BChl c | green sulfur bacteria | Frigaard, et.al., (2004) |
| BChl d | green sulfur bacteria | Frigaard, et.al., (2004) |
| BChl e | green sulfur bacteria | Frigaard, et.al., (2004) |
| phycoerythrin | unpublished absorption spectra from Govindjee's laboratory (Beckman DU spectrophotometer) | Govindjee (unpublished), and Ke (2001) |
| phycocyanin | unpublished absorption spectra from Govindjee's laboratory (Beckman DU spectrophotometer) | Govindjee (unpublished), and Ke (2001) |
| carotenoid | estimated *in vivo* absorption spectra in green algae.  NOTE: Type of carotenoid not specified. | Govindjee (1960) |
| Chl a flurorescence - | spinach chloroplasts | Govindjee and Yang, (1966) |





**Appendix A2.** Photosynthetic organisms reflectance data

| Organism type | Species measured | Instrument | Resolution | Source |
|---|---|---|---|---|
| Terrestrial vascular plants, temperate | Engelmann spruce lawn grass lodgepole pine maple oak piñon pine walnut | ASD FieldSpec | 3 nm (350-1000 nm) 10 nm (1000-2500 nm) | Clark, et al. (2003) USGS splib05a spectral database |
| Aquatic vascular plants – Calif. coast | *Phyllospadix torreyi* | ASD FieldSpec 350-2500P | 3 nm:.35-1 μm 10 nm:1-2.5 μm | Nancy Kiang |
| Aquatic vascular plants, seagrass | *Zostera capricorni Posidonia australis Halophila ovalis* | ASD FieldSpec FR | 3 nm:.35-1 μm 10 nm:1-2.5 μm | Fyfe, et al. (2003) |
| Lichens – temperate | *Acarospora Licedea Xanthoparmelia Xanthoria* | ASD FieldSpec | 3 nm:.35-1 μm 10 nm:1-2.5 μm | Clark, et al. (2003) USGS splib05a spectral database |
| Lichens – tropical | *Cladina skottsbergii Stereocaulon rocellodies* | ASD FieldSpec | 3 nm:.35-1 μm 10 nm:1-2.5 μm | Courtesy of Greg Asner |
| Moss – temperate | *Dicranum Plagiochila Polytrichum* | ASD FieldSpec | 3 nm:.35-1 μm 10 nm:1-2.5 μm | Clark, et al. (2003) USGS splib05a spectral database |
| Moss – Temperate | *Sphagnum capifollium Sphagnum cuspidatum Sphagnum pulchrum Sphagnum magellanicum Sphagnum papillosum* | ASD FieldSpec Pro | 1 nm: .35-2.5 μm | Harris, et al. (2005) |
| Algae – Calif. coast | Brown-*Macrocystis pyrifera* Red – unknown Green – *Ulva labata* | ASD FieldSpec 350-2500P | 3 nm:.35-1 μm 10 nm:1-2.5 μm | Nancy Kiang |
| Algae – snow | *Chlamydomonas nivalis* (Bauer) Wille | Ocean Optics S-2000, #754 | 0.66 nm @ 28-.86 μm | Gorton, et al. (2001) |
| Bacteria | Green filamentous Purple sulfur | ASD LabSpec VNIR-512 w/ optical fiber (LDG-GC 600/ 750, 25 m, Fujikuro, Tokyo, Japan) | 1.5 nm | Reinhard Bachofen (pers. comm.) Wiggli, et al. (1999) |
| Bacteria | *Rhodobacter sphaeroides Blastochloris viridis* | | 0.5 nm | Richard Cogdell, Andrew Gall (pers. comm.) |

Abbreviations:
ASD: Analytical Spectral Devices, Inc.
USGS: United State Geological Survey





## Appendix A3.  Spectral transmittance of light through algae in water in Figure 2b

In Figure 2b, the light incident at a depth of 5 cm in water with algae was calculated by estimating an absorbance coefficient from a measured reflectance spectrum, scaling this absorbance coefficient to approximate a density of algae, and then calculating the light transmitted through, given the surface incident radiation in Figure 2b.  The attenuated light, $I(z, \lambda)$, at a depth of z cm at wavelength $\lambda$, given incident light at the surface of $I(0, \lambda)$, spectral absorption coefficient of water $\alpha_{water}(\lambda)$ and of algae $\alpha_{algae}(\lambda)$ with a density parameter $\rho_{algae}$, is:

$$I(z, \lambda) = I(0, \lambda)e^{-\left(\alpha_{water} + \rho_{algae}\alpha_{algae}\right)z} \qquad (6)$$

For $\alpha_{water}(\lambda)$, we used water spectral absorption coefficient values as measured by the following sources:  200-380 nm, Segelstein (1981); 380-640 nm, Sogandares, et al. (1997);  640-2500 nm, Kou, et al., (1993).

For $\alpha_{algae}(\lambda)$, we took spectral reflectance measurements of fresh samples of a brown algae, *Macrocystis pyrifera* (kelp) with a FieldSpec Pro spectroradiometer (borrowed from Mike Eastwood and company at the JPL AVIRIS lab).  Since we could not measure transmittance directly, we estimated





transmittance from comparison to the behavior of a clover leaf modeled by the

PROSPECT leaf radiative transfer model of Jacquemoud and Baret (1990).  A

leaf transmittance spectrum is almost exactly the same as the reflectance, with the

sum of reflectance and transmittance scattering nearly all NIR near the red edge.

Therefore, we estimated the kelp transmittance from a scaling of the reflectance to

allow for about 5% absorbed NIR at the red edge, and then calculated the spectral

absorbance as 1- (reflectance + transmittance).  To approximate the density of the

algae in water, we simply let $\rho_{algae} = 0.10$.